\documentclass[3p,times,twocolumn,compress]{elsarticle}

\usepackage{ecrc}

\volume{00}

\firstpage{1}

\runauth{Q. Du et al.}

\jid{nppp}

\usepackage{amssymb}
\usepackage{amsmath}

\usepackage[figuresright]{rotating}
\usepackage{graphicx}
\usepackage[figurename=Fig.]{caption}
\usepackage{float}
\usepackage{tabu}
\usepackage[pdftex,colorlinks=true,linkcolor=blue,citecolor=blue,filecolor=blue,urlcolor=blue]{hyperref}
\usepackage[nameinlink]{cleveref}
\usepackage{subcaption}
\usepackage{ragged2e}
\usepackage{booktabs}
\usepackage[Symbol]{upgreek}
\usepackage{threeparttable}

\newcommand{\tabincell}[2]{\begin{tabular}{@{}#1@{}}#2\end{tabular}}
\crefformat{figure}{#2Fig.~#1#3}
\crefformat{equation}{#2Eq.~#1#3}
\begin{document}

\begin{frontmatter}



\dochead{}

\title{Measurement of the fast neutron background at the China Jinping Underground Laboratory}

\author[label1,label2]{Q. Du\fnref{cor1}}
\fntext[cor1]{Main contributor}
\author[label1]{S.T. Lin\corref{cor2}}
\cortext[cor2]{Corresponding author: stlin@scu.edu.cn}
\author[label1]{S.K. Liu}
\author[label1]{C.J. Tang}
\author[label3]{L. Wang}
\author[label2]{W.W. Wei}
\author[label4]{H.T. Wong}
\author[label1]{H.Y. Xing}
\author[label3]{Q. Yue}
\author[label2]{J.J. Zhu}

\address[label1]{College of Physical Science and Technology, Sichuan University, Chengdu 610064}
\address[label2]{Institute of Nuclear Science and Technology, Sichuan University, Chengdu 610064}
\address[label3]{Key Laboratory of Particle and Radiation Imaging (Ministry of Education) and Department of Engineering Physics, Tsinghua University, Beijing 100084}
\address[label4]{Institute of Physics, Academia Sinica, Taipei 11529}   

\begin{abstract}
We report on the measurements of the fluxes and spectra of the environmental fast neutron background at the China Jinping Underground Laboratory (CJPL) with a rock overburden of about 6700 meters water equivalent, using a liquid scintillator detector doped with 0.5\% gadolinium. 
The signature of a prompt nuclear recoil followed by a delayed high energy $\upgamma$-ray cascade is used to identify neutron events. The large energy deposition of the delayed $\upgamma$-rays from the $(n, \upgamma)$ reaction on gadolinium, together with the excellent n-$\upgamma$ discrimination capability provides a powerful background suppression which allows the measurement of a low intensity neutron flux. The neutron flux of $(1.51\pm0.03\,(stat.)\pm0.10\,(syst.))\times10^{-7}$\,cm$^{-2}$s$^{-1}$ in the energy range of 1\,--\,10\,MeV in the Hall~A of CJPL was measured based on 356 days of data. In the same energy region, measurement with the same detector placed in a one meter thick polyethylene room gives a significantly lower flux of $(4.9\pm0.9\,(stat.)\pm0.5\,(syst.) )\times10^{-9}$~cm$^{-2}$s$^{-1}$ with 174 days of data. This represents a measurement of the lowest environmental fast neutron background among the underground laboratories in the world, prior to additional experiment-specific attenuation. Additionally, the fast neutron spectra both in the Hall~A and the polyethylene room were reconstructed with the help of GEANT4 simulation.
\end{abstract}


\end{frontmatter}


\section{Introduction}
Deep underground sites provide a unique opportunity to explore rare-event phenomena including the direct searches for dark matter~\cite{dark-matter}, proton decay~\cite{proton-decay}, neutrinoless double beta decay ($0\nu\beta\beta$)~\cite{neutrinoless}, neutrino oscillation experiments~\cite{neutrino}, and so on.
A comprehensive range of underground experiments is sensitive to the neutron and its induced background. 

\begin{figure*}
\centering
\begin{subfigure}{\columnwidth}
\includegraphics[width=\linewidth, height=4.82 cm]{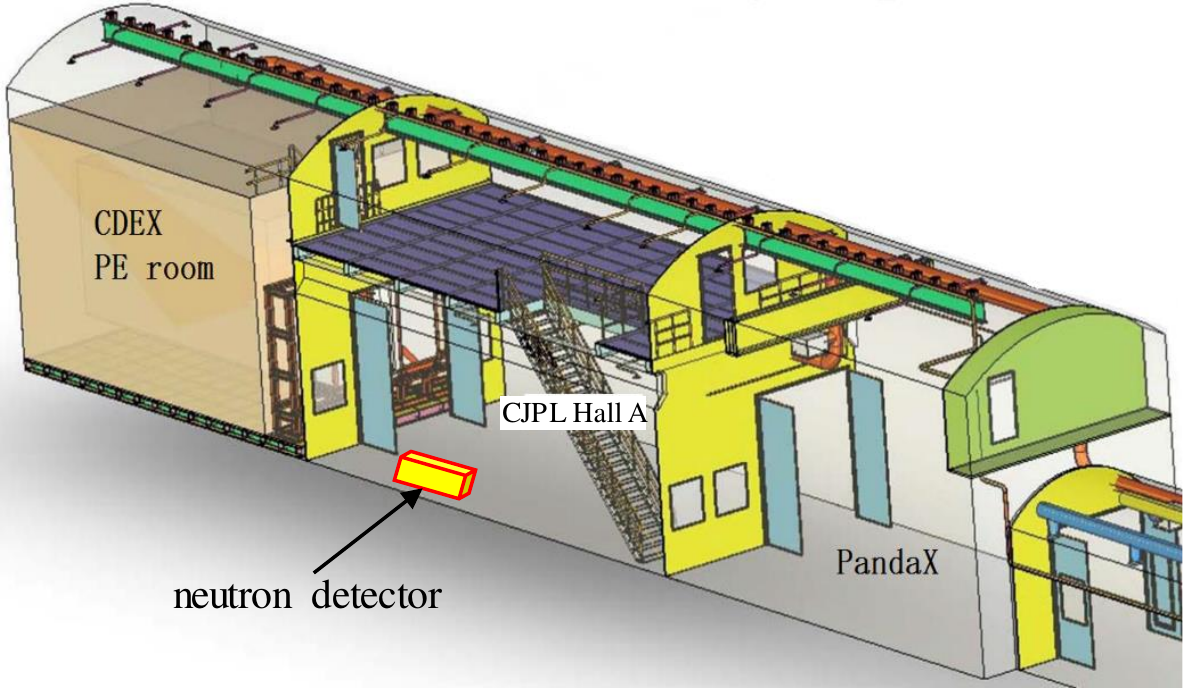}
\caption{\label{fig:CJPL_hall}}
\end{subfigure}
\hfill
\begin{subfigure}{\columnwidth}
\includegraphics[width=\linewidth, height=4.82 cm]{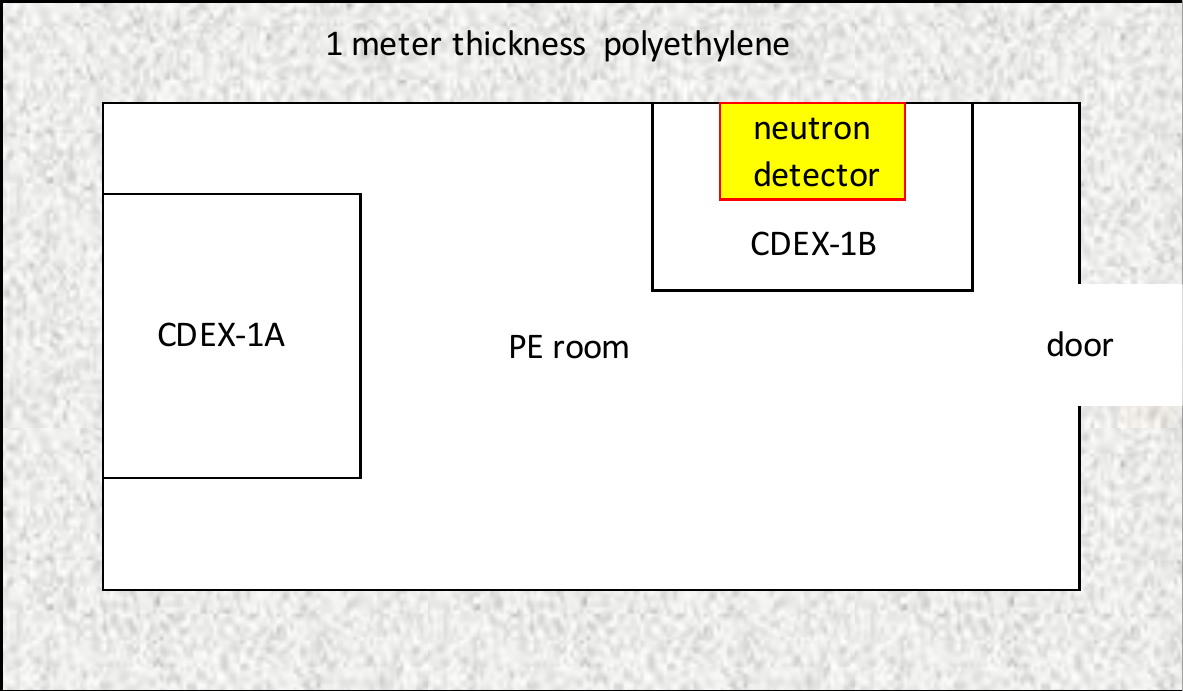}
\caption{\label{fig:CJPL_PE}}
\end{subfigure}
\caption{\label{fig:CJPL}(color online) The structure of the China Jinping Underground Laboratory and the location of the neutron detector. (a) The position of the Hall A and the polyethylene (PE) room; (b) The structure of the PE room.}
\end{figure*}

In addition to the highly suppressed cosmic-ray induced neutrons, the majority of neutrons at the deep underground site is produced in the rock through the spontaneous fission of $^{238}$U and the ($\alpha$, n) reactions of light nuclei bombarded by the $\alpha$-particles emitted in the U/Th decay chains.
An additional neutron background is taken along with the infrastructure and the experimental setup of laboratories. 
Nuclear recoils and other interactions due to the environmental neutron background can restrict experimental sensitivities or bring false positive signals in the studies of rare phenomena. 
Therefore, the understanding of neutron spectrum and identifying neutron sources are substantial issues for background reduction and guide the design of shielding systems for the next generation of large-scale experiments.

The China Jinping Underground Laboratory (CJPL)~\cite{CJPL} with about 2400\,m (6700\,m water equivalent) rock overburden is the deepest operating underground laboratory, located in Sichuan, China.
The cosmic-ray flux at CJPL is measured to be $(2.0\pm0.4)\times10^{-10}$\,cm$^{-2}$s$^{-1}$~\cite{Muon}. 
Low background counting facilities are installed and implemented. The environmental gamma radioactivity at CJPL has been studied. The contamination levels of U and Th in the rock at CJPL are demonstrated to be one order of magnitude lower than those at the surface due to the carbonate components (96\% calcite and $<$4\% micritic limestone)~\cite{Gamma}.
The science program of dark matter experiments at CJPL, in the first phase, includes CDEX (the China Dark matter EXperiment)~\cite{CDEX} carried out in a one meter thick polyethylene (PE) room, and PandaX (the Particle and Astrophysical Xenon Detector)~\cite{PandaX} conducted in the hall. The layout of CJPL with the operated experiments is depicted in \cref{fig:CJPL}.

The thermal neutron flux was measured with a gaseous $^{3}$He proportional ionization chamber in the Hall~A of CJPL, giving a flux of $(4.00\pm0.08)\times10^{-6}$\,cm$^{-2}$s$^{-1}$~\cite{thermal neutron}. The same measurement was performed inside the PE room. A preliminary thermal neutron flux of $(3.18\pm0.97)\times10^{-8}$\,cm$^{-2}$s$^{-1}$ was derived~\cite{thermal neutron PE}. A Bonner multi-sphere neutron spectrometer, with $^{3}$He ionization chambers inside the polyethylene spheres with different thicknesses, has been used to investigate the neutron background below 20\,MeV in the Hall~A and gives a overall neutron flux of $(2.69\pm1.02)\times10^{-5}$\,cm$^{-2}$s$^{-1}$~\cite{Hu Q}. Due to the limited detector sensitivity and resolution, the fast neutron background inside the PE room has not been reported.

This work presents the background measurement of fast neutron fluxes and spectra both in the Hall A and inside the PE room at CJPL, using a 28\,liter liquid scintillator detector doped with 0.5\% gadolinium. 
The location of the neutron detector at CJPL is also shown in \cref{fig:CJPL}.
The cascade of prompt and delayed signatures offers a powerful background suppression against the alpha-contamination from the liquid scintillator. 
The detector configuration and the data acquisition system is described in \Cref{sec:detector}. The energy calibration, the energy resolution of the detector and the efficiency of signal selection are derived by comparison with simulations and are presented in \Cref{sec:calibration}. Following the data analysis in \Cref{sec:data analysis}, the results and the conclusions are reported in the last two sections.

\section{Experimental setup}
\label{sec:detector}

The (n, $\upgamma$) reaction on gadolinium in the liquid scintillator allows the discrimination of neutron events from the background with the prompt-delayed time coincidence method. This method is based on the time delay between the prompt nuclear recoil signal and the $\upgamma$-cascade produced by the neutron capture of the thermalized neutrons. Nuclear recoils due to the multiple elastic scatterings of fast neutrons constitute the prompt signal with signal profiles satisfying the pulse shape criteria. 
After thermalized, the neutrons diffuse in the detector for a few microseconds ($\sim 7$\,$\upmu$s) before they are captured on the gadolinium and soon after, emit a $\upgamma$-cascade giving rise to the delayed signal. This delayed coincident signature is different from the decay sequences of the $\beta$-$\alpha$ or the $\alpha$-$\alpha$ cascade decays from U and Th series. Therefore, the gadolinium doped liquid scintillator (Gd-LS) neutron detector is insensitive to its intrinsic U/Th contamination. 

\begin{figure}[t]
\includegraphics[width=\linewidth]{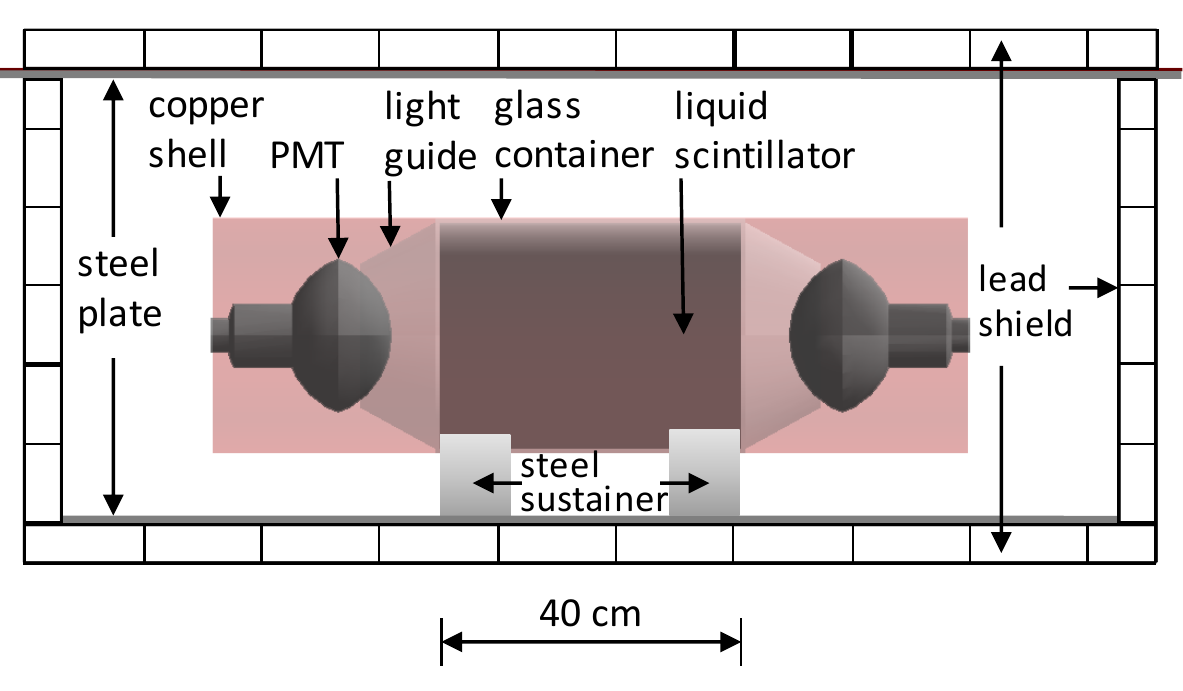}
\caption{\label{fig:detector}(color online) The schematic diagram of the liquid scintillator detector. The liquid scintillator is EJ-335. The container for EJ-335 is a quartz glass cylinder with 30\,cm diameter and 40\,cm length. The copper shell is used for support and shielding. The detector is located in a lead castle with 5\,cm thickness. The top steel plate is used to hold the roof of the lead castle. The bottom one is used to support the detector with two steel sustainers.}
\end{figure}

\begin{figure}
\includegraphics[width=\linewidth]{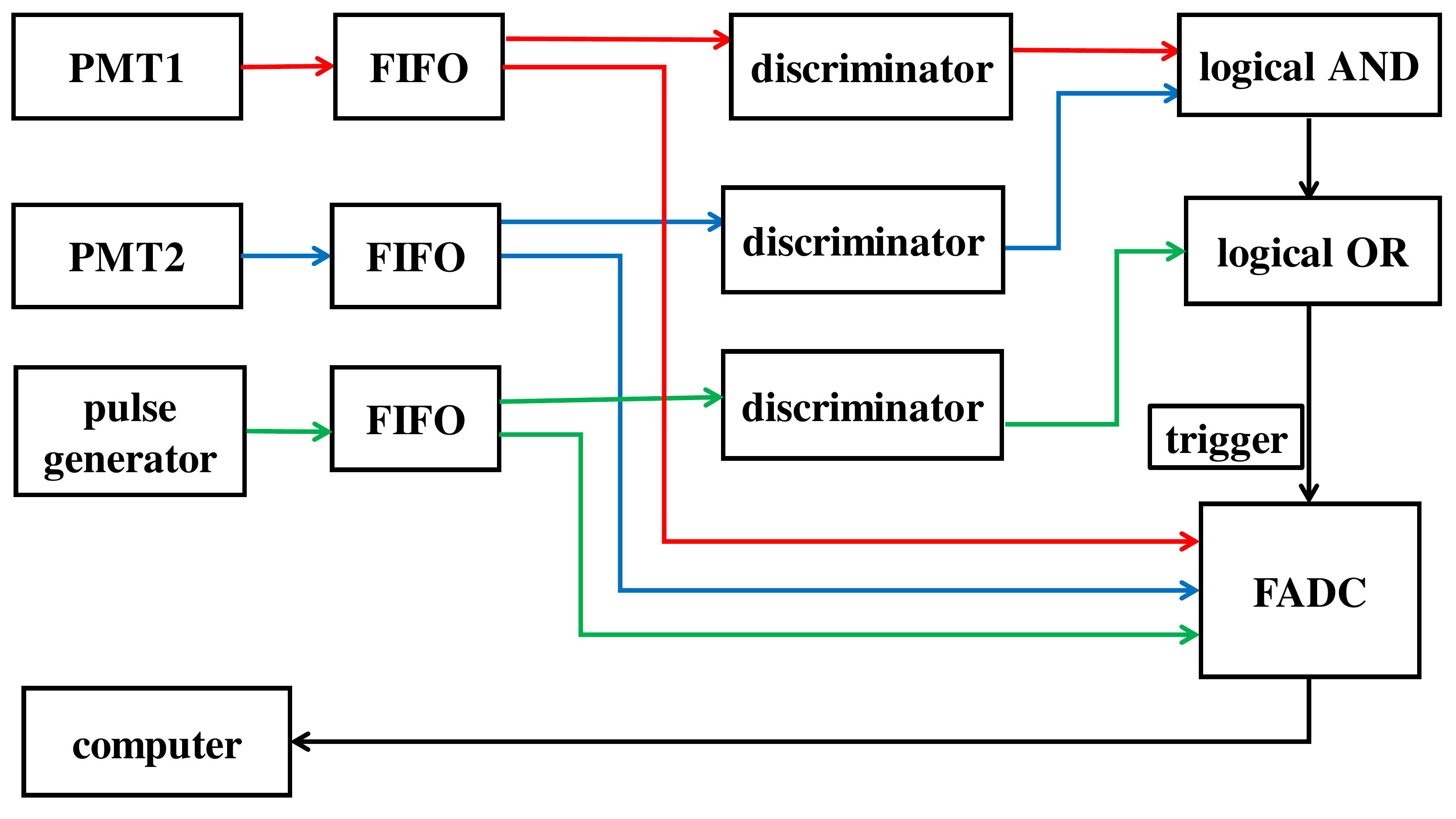}
\caption{\label{fig:DAQ}(color online) The flow scheme of the data acquisition (DAQ) system. The duplicate signals are generated by linear fan-in/fan-out (FIFO) modules from the two PMTs, as well as the periodic pulse generator. One is fed into the discriminator, and the other is read out by a flash analog-to-digital converter (FADC). The DAQ triggers are provided by coincident signals of the two PMTs and by random events produced by pulse generator.}
\end{figure}

The Gd-LS detector has been used for many years by several experimental groups for measurements of the neutron background at underground laboratories, such as Boulby~\cite{Boulby} and Aberdeen Tunnel~\cite{Aberdeen Tunnel}, or for neutrino experiments, such as Double Chooz~\cite{Double Chooz}, RENO~\cite{RENO} and Daya Bay~\cite{Daya Bay} measuring the neutrino mixing angle $\theta_{13}$.
The employed Gd-LS in this work is of the type EJ-335 produced by Eljen Technology Company and is an organic scintillator loaded with 0.5\% gadolinium by weight~\cite{EJ-335}.

The Gd-LS is filled in a cylindrical container with 30\,cm diameter and 40\,cm length made of quartz glass that is wrapped with PTFE sheet for high diffuse reflection.
The two flat surfaces of the glass vessel are optically connected by light guides to two 8\,inch photomultiplier tubes (PMTs) from Hamamatsu (type R5912-02).
The neutron detector is supported by a 3\,mm thick oxygen-free copper cylinder, placed in a 5\,cm thick lead castle with the dimensions of $150\times60\times70$\,cm$^{3}$ depicted in \cref{fig:detector}. A hole was made available to insert radioactive sources into the lead castle to perform calibrations. 

The flow scheme for data acquisition (DAQ) system is illustrated in \cref{fig:DAQ}. The signals are duplicated by linear fan-in/fan-out (FIFO) modules from the two PMTs, as well as from the periodic pulse generator with high accuracy used as random trigger (RT) which allows the measurement of live time and selection efficiency. One signal is fed into the discriminator, and the other is read out by a flash analog-to-digital converter (FADC) with 500\,MHz sampling rate and 8\,bit resolution. The triggers for the DAQ are provided by the coincident signals from the two PMTs and by the RT. 

\section{Calibration measurements}
\label{sec:calibration}
\subsection{Energy calibration}
\label{}
\begin{figure}
\includegraphics[width=\linewidth]{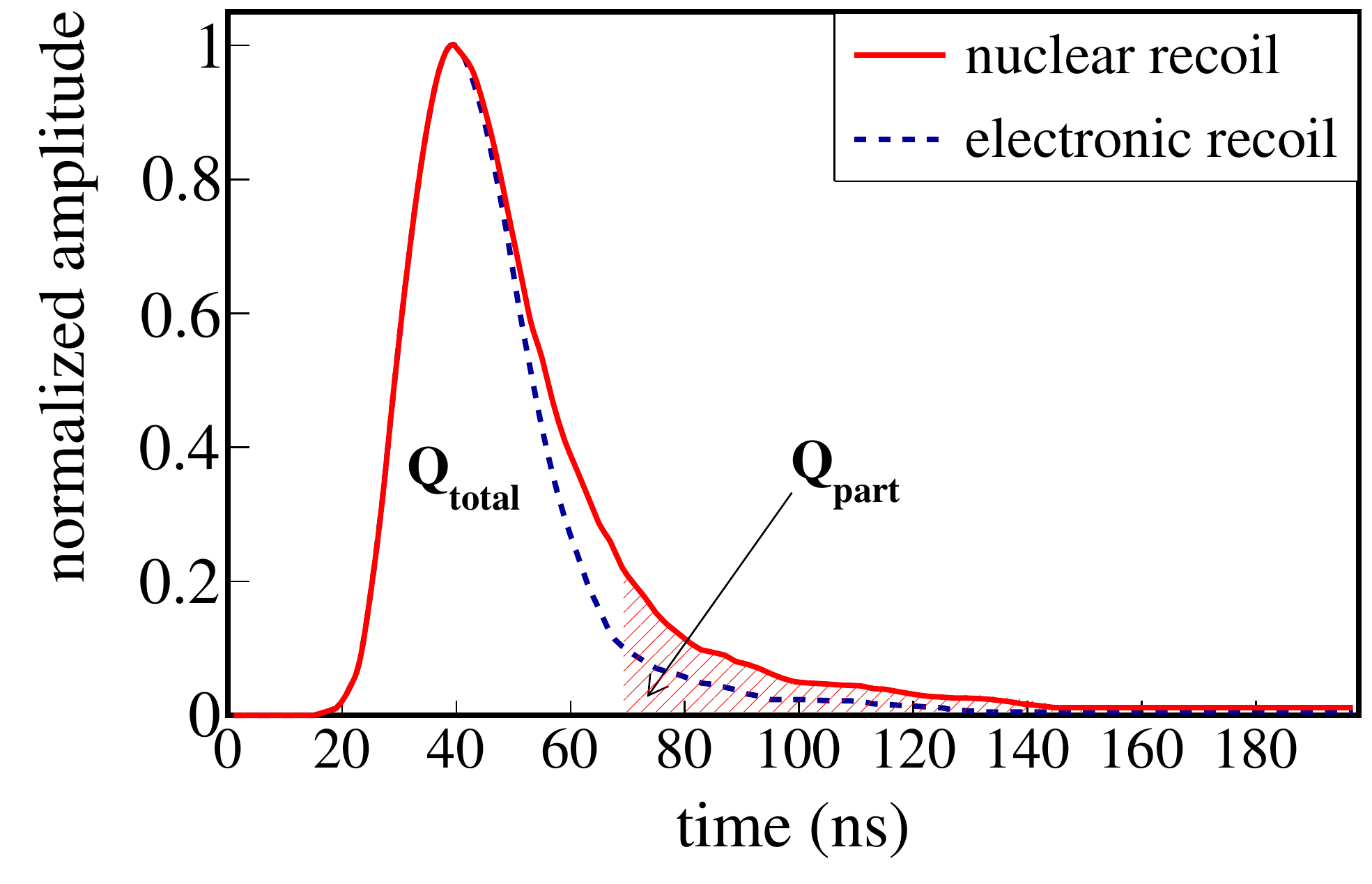}
\caption{\label{fig:parameter}(color online) The average waveform of nuclear recoils and electronic recoils. `$Q_{total}$' is the total area of the waveform which is proportional to the charge. `$Q_{part}$' is the area of the tail of the waveform. The discrimination factor ($Dis$) is defined as `$Q_{part}/Q_{total}$'.}
\end{figure}

The shape of the average waveform both for nuclear recoils and electronic recoils are displayed in \cref{fig:parameter}. `$Q_{total}$' and `$Q_{part}$' denote the total and the partial integration of the pulse within (-40, 160)\,ns and (30, 160)\,ns time window with respect to the instant of the maximum height, respectively. A location independent energy of a signal can be defined as~\cite{Mei_detector}

\begin{equation}
\label{eq:QEnergy}
 E = a \cdot \sqrt{Q_{total_{1}} \cdot Q_{total_{2}}}
\end{equation}
where the subscripts `$1$' and `$2$' represent the two PMTs respectively, and `$a$' is a constant, evaluated from measurements with calibration sources.

\begin{figure*}
\centering
\begin{subfigure}{0.32\textwidth}
\includegraphics[width=\textwidth]{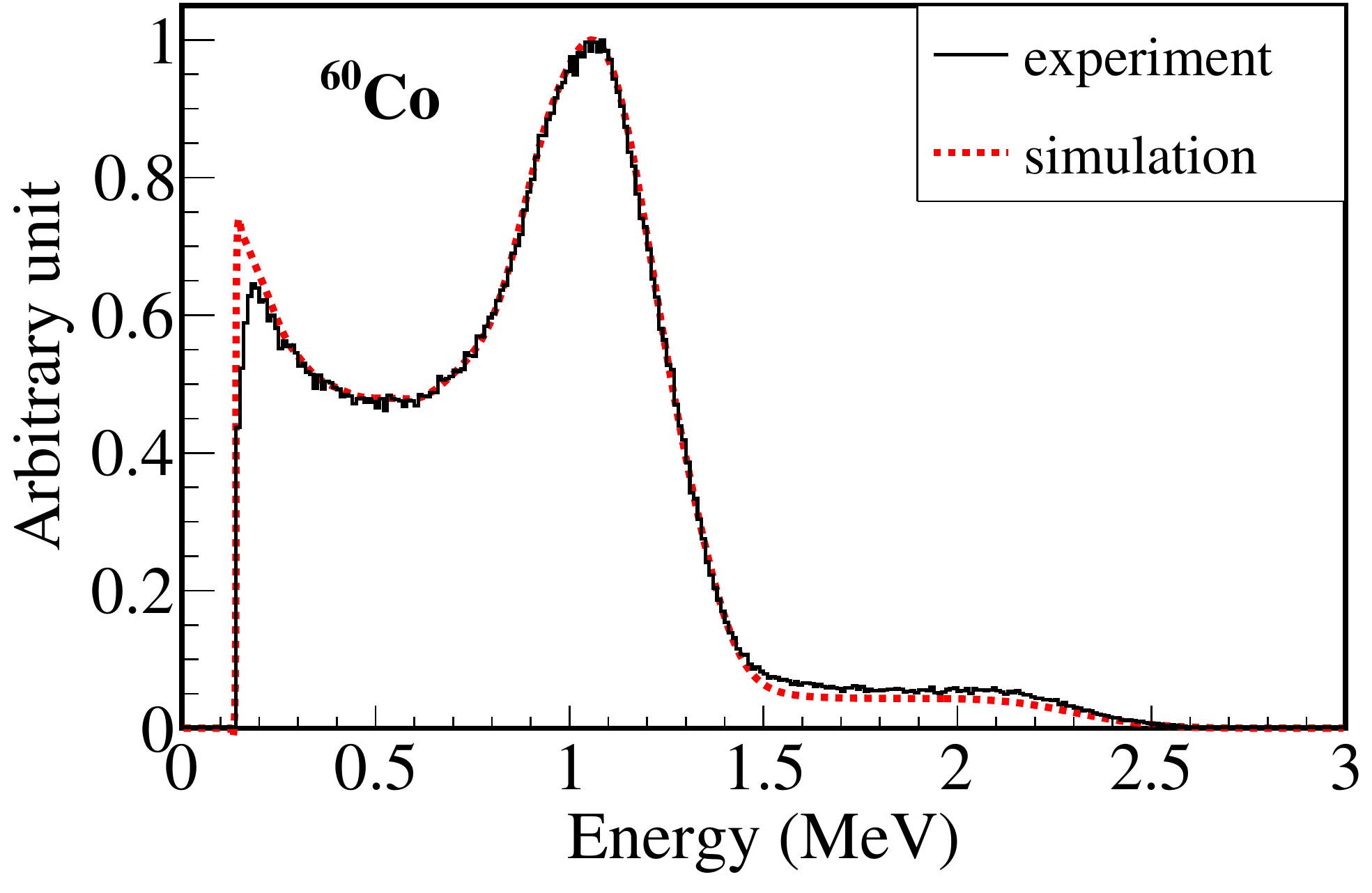}
\caption{\label{fig:Co60} }
\end{subfigure}
\hfill
\begin{subfigure}{0.32\textwidth}
\includegraphics[width=\textwidth]{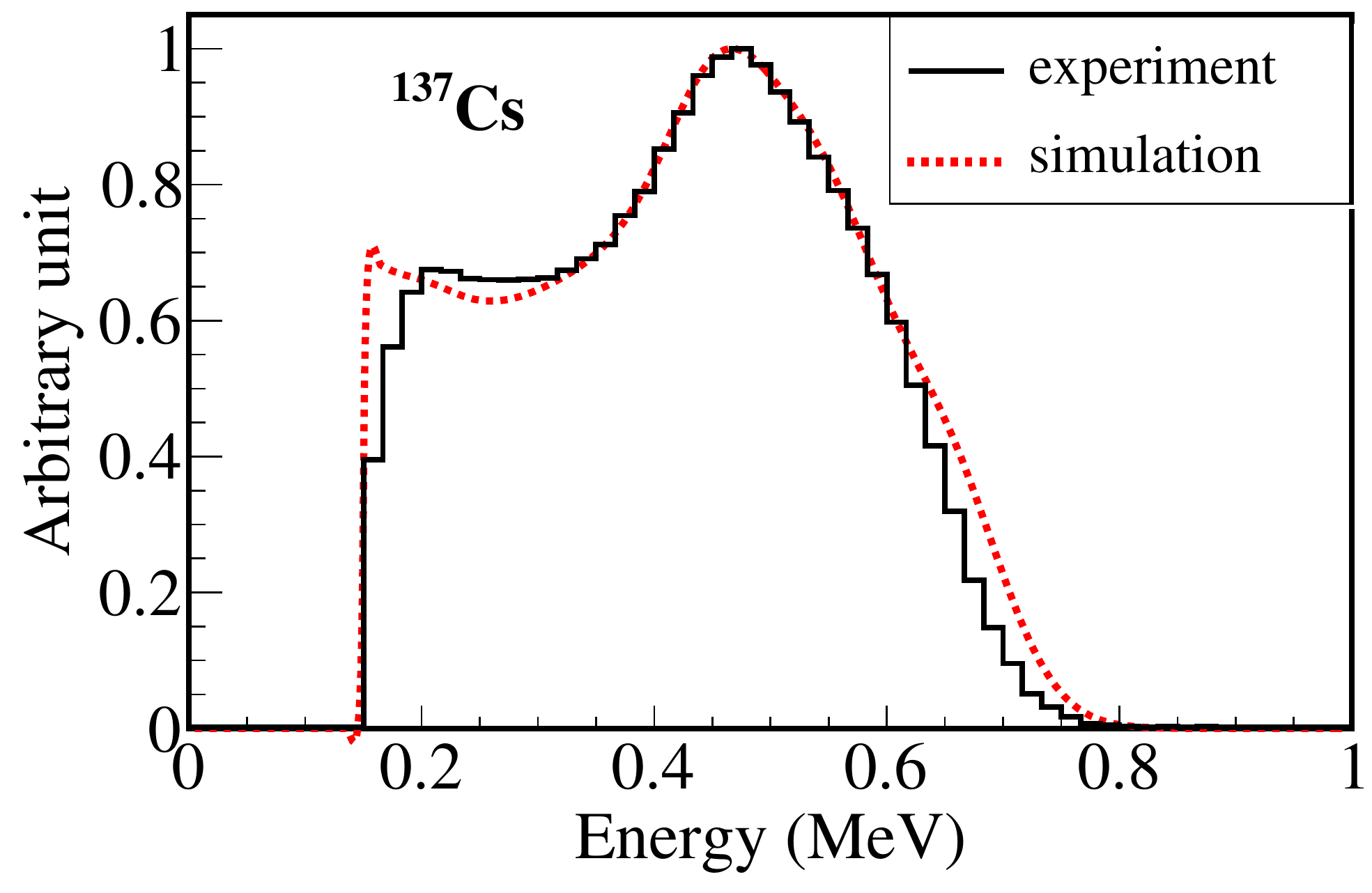}
\caption{\label{fig:Cs137} }
\end{subfigure}
\hfill
\begin{subfigure}{0.32\textwidth}
\includegraphics[width=\textwidth]{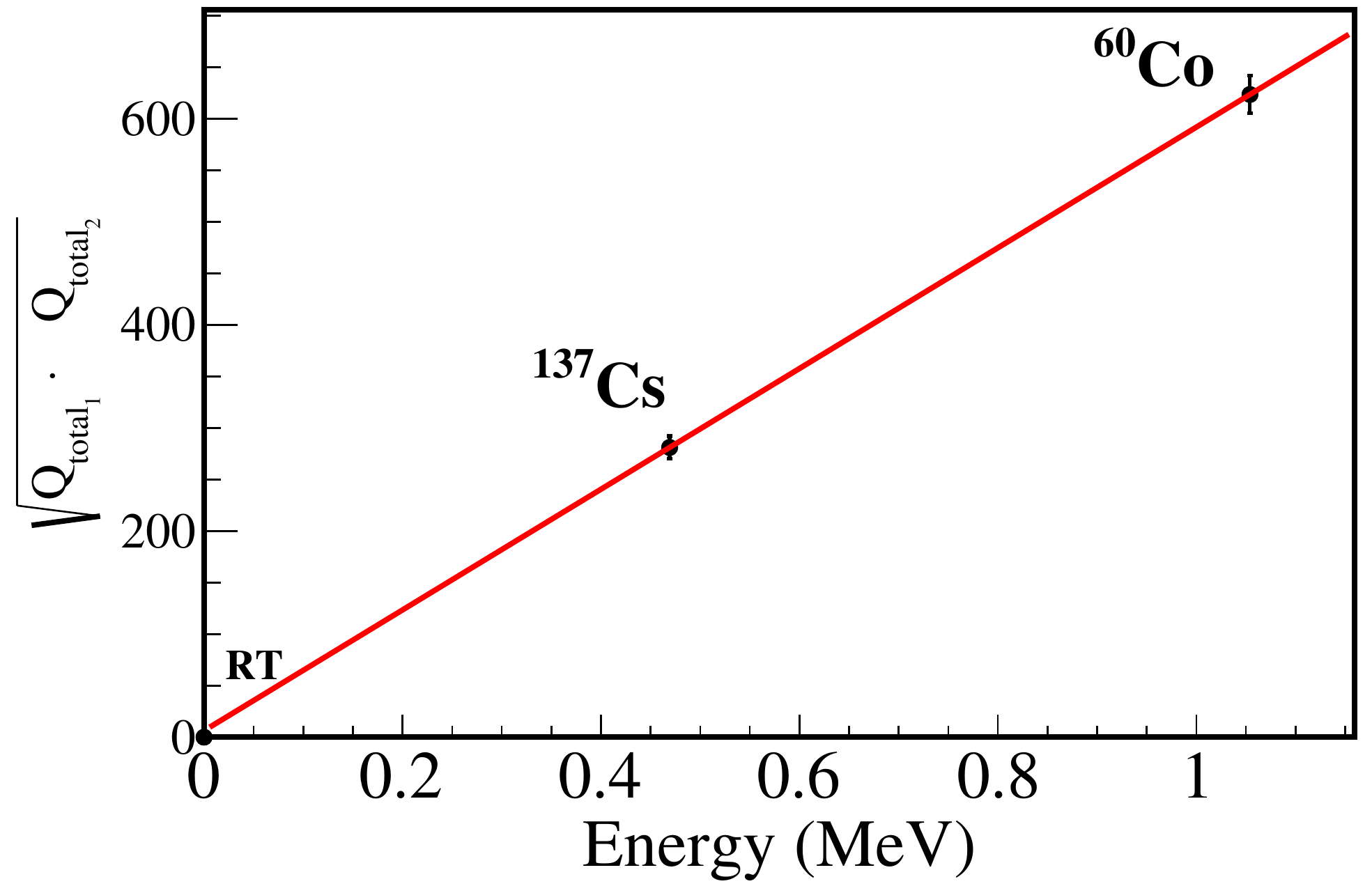}
\caption{\label{fig:calibration}}
\end{subfigure}
\caption{\label{fig:EnergyCalibration}(color online) 
Energy calibration with $\gamma$-ray sources. (a) and (b) show the energy spectra of $^{60}$Co and $^{137}$Cs, respectively, where the simulated spectra are smeared out with a resolution function given by $(\sigma/E)^2=\alpha+\beta/E$ where $\alpha$ = 0.03 and $\beta$ = 3.5\,keV; (c) shows the linear calibration function.}
\end{figure*}

Two gamma sources, $^{137}$Cs and $^{60}$Co were inserted in the lead castle for two independent energy calibrations. The characteristics of the Compton edges were estimated via full GEANT4 simulations including detector resolution smearing. The spectra from the simulation and the calibration data are compared in \cref{fig:Co60} for the $^{60}$Co and \cref{fig:Cs137} for the $^{137}$Cs source. The pedestal zero energy was determined by the RT. An excellent linearity is observed, as shown in \cref{fig:calibration}. The energy resolution function is given by $(\sigma/E)^2=\alpha+\beta/E$~\cite{Boulby}, where the parameters $\alpha$ = 0.03 and $\beta$ = 3.5\,keV were obtained by matching with the simulations.

\subsection{Efficiency calibration by neutron source}
\label{sec:efficiency}

An AmBe neutron source was deployed to provide the calibration of the neutron detection efficiency. 
It was placed at 4.3\,m external to the lead shield.
Both the prompt-delayed time coincidence (\Cref{sec:detector}) and pulse shape discrimination (PSD) methods have been used to select the neutron-induced nuclear recoils (hereafter denoted as nuclear recoils) in a high $\upgamma$-ray background environment. The PSD method depends on the difference of the tail of the waveform between nuclear recoil (neutron or alpha induced) and electronic recoil ($\upgamma$-ray induced) (see \cref{fig:parameter}). The discrimination factor ($Dis$) is defined to differentiate nuclear recoils and $\upgamma$-ray events, defined as

\begin{equation}
\label{eq:Dis}
Dis=\frac{Q_{part_{1}} + Q_{part_{2}}}{Q_{total_{1}} + Q_{total_{2}}}
\end{equation}
where the subscripts of `$1$' and `$2$' represent the two PMTs respectively.

\begin{figure*}
\begin{subfigure}{0.32\textwidth}
\includegraphics[width=\textwidth]{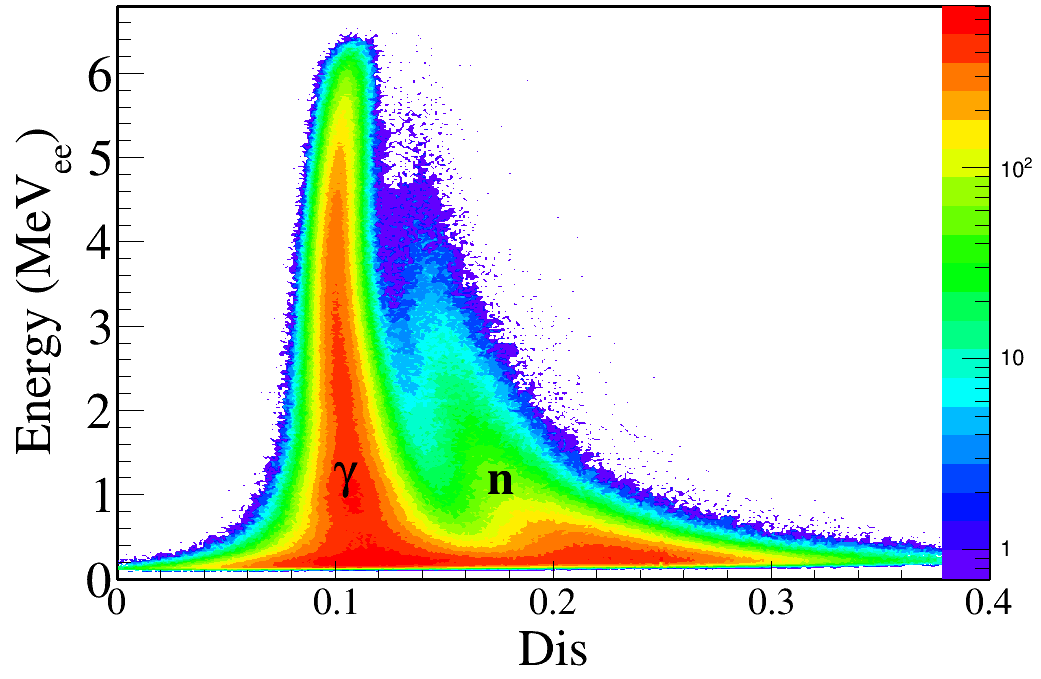}
\caption{\label{fig:AmBe}}
\end{subfigure}
\hfill
\begin{subfigure}{0.32\textwidth}
\includegraphics[width=\textwidth]{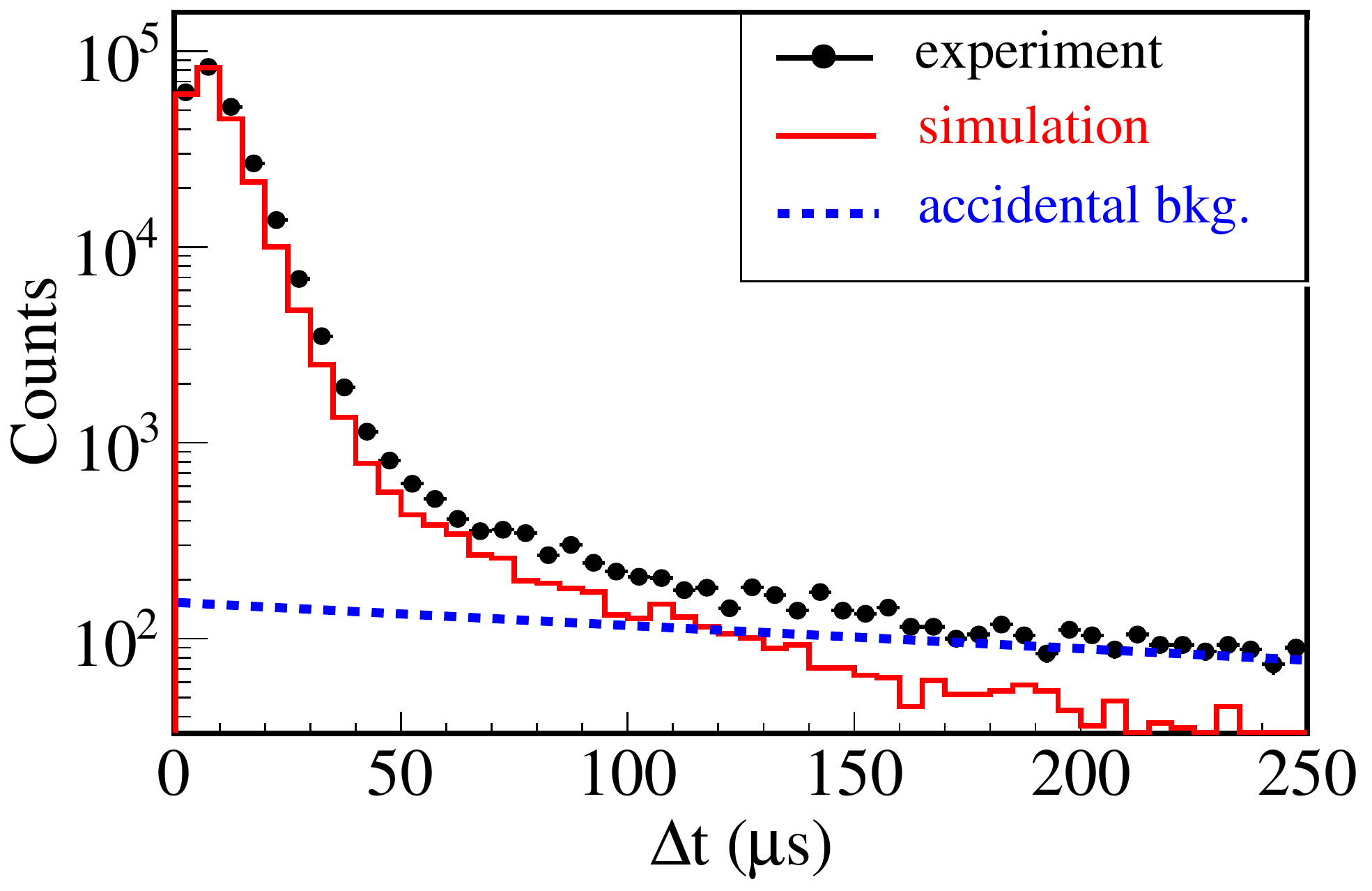}
\caption{\label{fig:AmBeTimeInterval}}
\end{subfigure}
\hfill
\begin{subfigure}{0.32\textwidth}
\includegraphics[width=\textwidth]{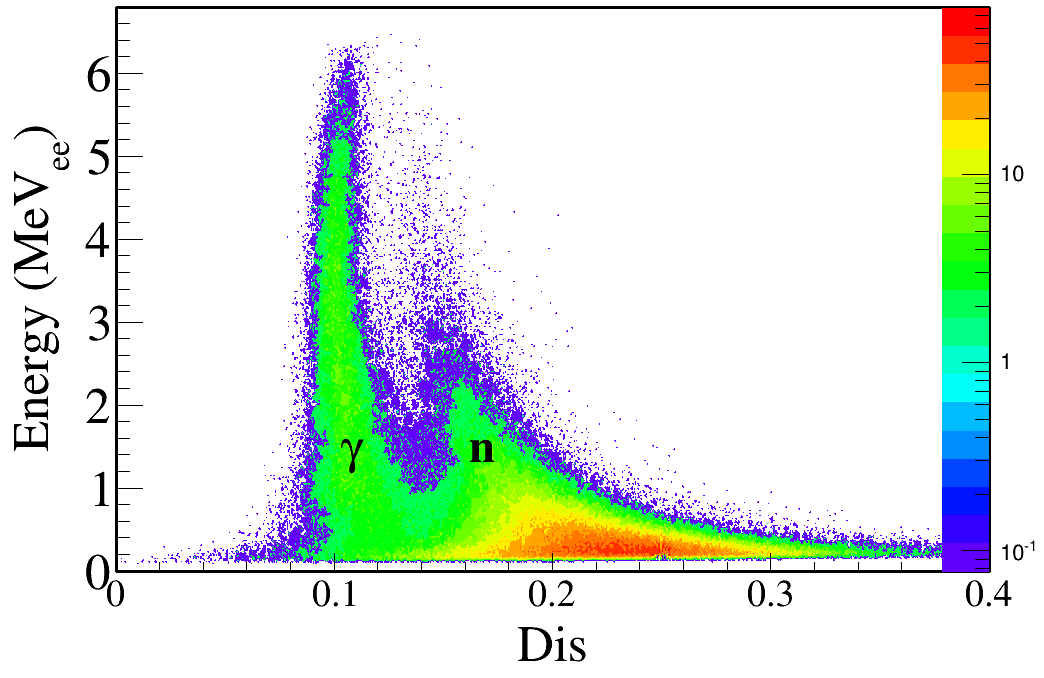}
\caption{\label{fig:AmBeNeutron}}
\end{subfigure}
\caption{\label{fig:AmBe_selection}(color online) The selection procedure of the prompt nuclear recoils from the neutron source data. (a) The distribution of the discrimination factor ($Dis$) at different energies both for the nuclear recoils (n) and the $\upgamma$-ray events; (b) The measured time interval distribution of the selected prompt-delayed signals together with the simulated ones and the accidental background; (c) The selected prompt events following the energy and time window conditions (see text for details).}
\end{figure*}

\cref{fig:AmBe_selection} illustrates the selection procedure of the prompt nuclear recoils from neutron source data. The discrimination factors $Dis$ at different energies for both nuclear recoils and $\upgamma$-ray events are depicted in \cref{fig:AmBe}. Their $Dis$ distributions are well-separated. The selection procedure is described as follows:

(1). $\upgamma$-rays from neutron capture were selected from the data (displayed in \cref{fig:AmBe}) with a minimum energy of 3\,MeV as the delayed signals. This condition suppressed most of the environmental $\upgamma$-ray background from natural radioactivity, in particular, the 2.6\,MeV $\upgamma$-rays from $^{208}$Tl. The residual nuclear recoils in this region were rejected by a PSD cut at $Dis<0.11$. The efficiency of this cut was obtained by fitting the $Dis$ with two Gaussian functions.

(2). The prompt signal ahead of the delayed signal from step (1) was regarded as the nuclear recoil. The software threshold for the prompt nuclear recoils was set to 0.23\,MeV$_{ee}$ (equivalent electron energy), corresponding to 1\,MeV neutron deposited energy due to the quenching factor of the Gd-LS.

(3). \cref{fig:AmBeTimeInterval} shows the time interval distribution between the prompt and delayed signals, compared to GEANT4 simulation, as well as the accidental background events. A coincidence time window of (2, 30)\,$\upmu$s was required to exclude most of the accidental background. The selected prompt events are shown in \cref{fig:AmBeNeutron}.

(4). There is residual coincident $\upgamma$-ray background which survive the selection, as shown at the left band of \cref{fig:AmBeNeutron}. To extract the number of prompt nuclear recoils as well as the nuclear recoil spectrum, double-Gaussian is used to fit the $Dis$ distribution for each energy bin with step of 0.1\,MeV$_{ee}$.

\begin{figure}
\includegraphics[width=\linewidth]{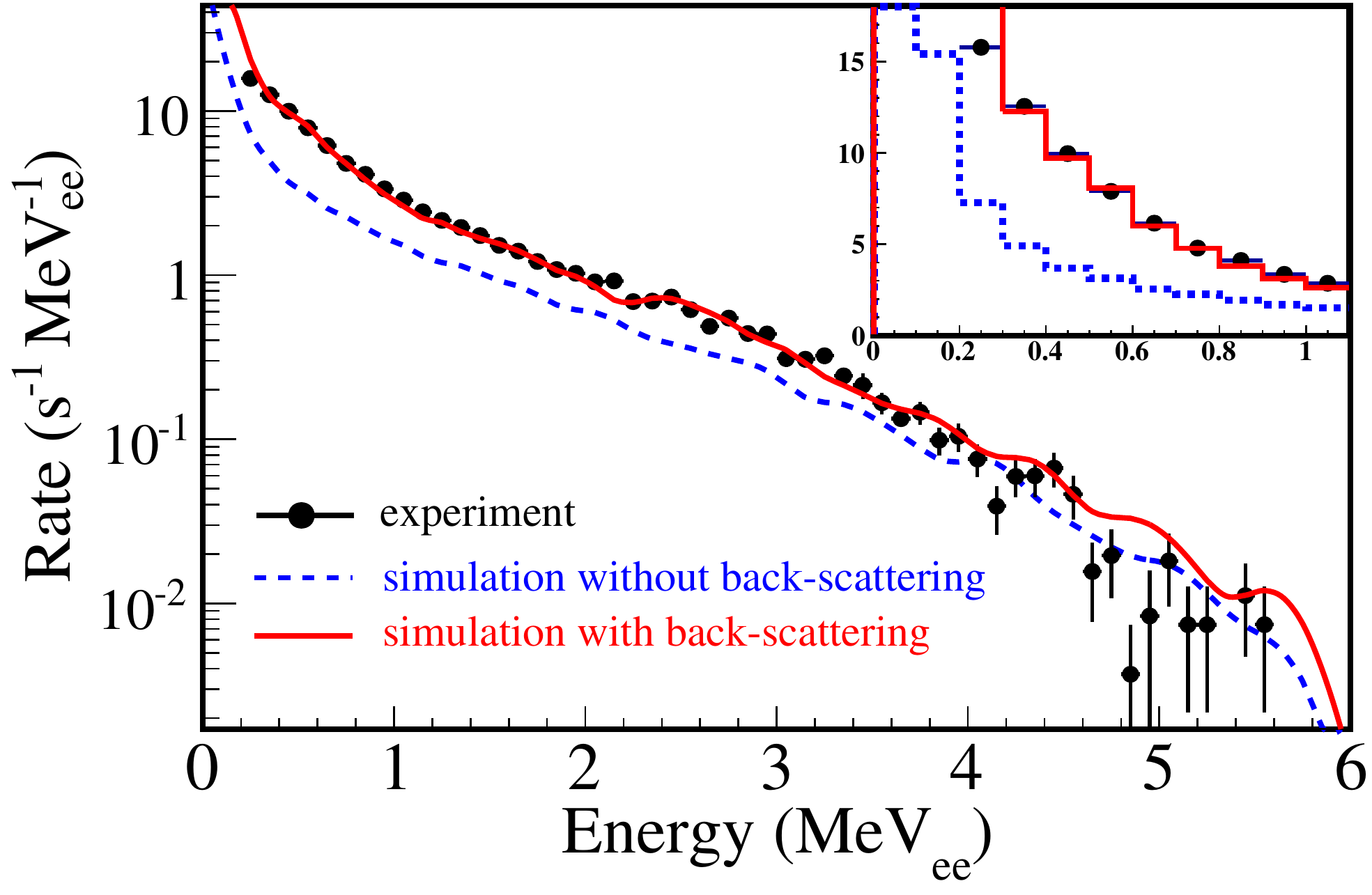}
\caption{\label{fig:AmBeRecoilSpectrum}(color online) The nuclear recoil spectrum of the AmBe neutron source. The solid black dots is the measured recoil spectrum after the subtraction of coincident $\upgamma$-ray background. The dashed blue curve corresponds to the simulated recoil spectrum without the back-scattering from the room. The solid red curve is the one when the back-scattering was taken into account. The simulated spectra are normalized to the neutron rate of the AmBe source and therefore absolute values can be compared with the experiment.}
\end{figure}

\cref{fig:AmBeRecoilSpectrum} is the final measured nuclear recoil spectrum of the AmBe neutron source after the subtraction of the coincident $\upgamma$-ray background. GEANT4 (version 4.9.6.p04 with QGSP\_BERT\_HP module physics list) was used to simulate the nuclear recoil spectrum and compare to the experiment. The selection of the nuclear recoils in simulation followed a similar procedure as for the experimental data. A standard neutron spectrum of ISO-8529 AmBe was adopted as the input for the simulation. In \cref{fig:AmBeRecoilSpectrum}, the dashed curve is the simulated nuclear recoil spectrum without the back-scattering from the room. Discrepancies can been seen between the experiment and the simulation, especially in the low energy region. After taking into account the room back-scattering in the simulation, the solid curve gives an excellent agreement when compared to the measured data. 

The neutron detection efficiency is defined as the ratio of the number of selected nuclear recoils to the number of neutrons entering the Gd-LS detector with the incident energy larger than 1\,MeV. The rate of the emitted neutrons from the AmBe source is 100080\,s$^{-1}$ with 3.5\% accuracy. The relative effective solid angle can be calculated based on simulation. Finally, the detection efficiency is measured to be $(11.13\pm 0.42)\%$ under the selections mentioned above with 1\,MeV neutron energy threshold. The uncertainties are mainly due to the statistical uncertainty given by the double-Gaussian fit and the systematic uncertainty from the neutron source activity. The simulated efficiency is $(10.52\pm 0.12)\%$, which is consistent with the measured result.

\subsection{Neutron spectrum reconstruction}
\label{sec:AmBe spectrum unfolding}
The physical quantity of interest is the fast neutron energy spectrum at CJPL.
This can be derived from the measured nuclear recoil spectrum with the SAND-II method~\cite{SAND}:

\begin{equation}
\label{eq:SAND-II}
\begin{split}
\Phi^{j+1}_{i} = \Phi^{j}_{i} exp \left( {\frac{ \sum_{k=1}^{K} W^{j}_{ik} ln (U_{k}^{j}) }{ \sum_{k=1}^{K} W^{j}_{ik} } }\right)\\
j=0,1,\cdots,J\\
\end{split}
\end{equation}
\begin{equation*}
W_{ik}^{j} = \frac{R_{ki}\Phi_{i}{j}}{\sum_{i=1}^{I}R_{ki}\Phi_{i}^{j}}\frac{N_{k}^{2}}{\sigma_{k}^{2}},\quad
U_{k}^{j} = \frac{N_{k}}{\sum_{i=1}^{I} R_{ki} \Phi_{i}^{j} \Delta E_{i}}
\end{equation*}
where
$R_{ki}$ is the detector response function of the neutron in the $k^{th}$ energy bin of the recoil spectrum;
$N_{k}$ is the measured nuclear recoil in the $k^{th}$ energy bin and, $\sigma_{k}$ corresponds to its uncertainty; $K$ is the maximum value for $k$;
 $\Phi_{i}^{j}$ is the $j^{th}$ iteration result at the $i^{th}$ energy bin of incident neutron spectrum and, $\Delta E_{i}$ corresponds to its bin width. 
An initial spectrum ($\Phi_{i}^{0}$) with prior information should be provided to initiate the iterations. The maximum number of the iterations is denoted by J such that $\Phi_{i}^{J}$ is the final unfolded neutron spectrum.

\begin{figure}
\includegraphics[width=\linewidth]{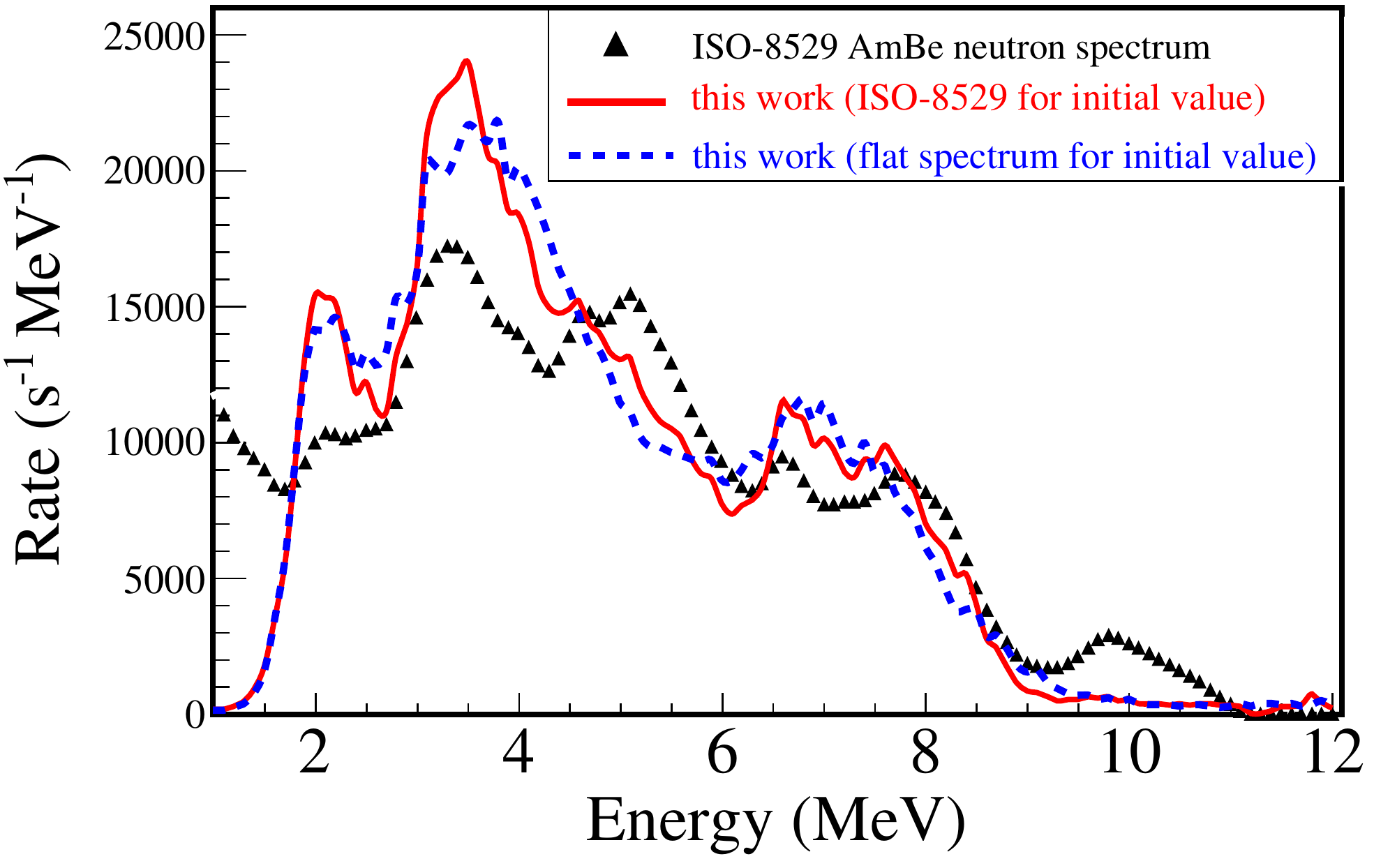}
\caption{\label{fig:AmBeUnfoldingSpectrum}(color online) The unfolded neutron spectrum of the AmBe source. Two different initial spectra were fed into the unfolding program. The solid red curve is the unfolding result using the neutron spectrum of ISO-8529 AmBe as the initial value. The dashed blue curve is the unfolding result with a flat spectrum as the initial value. The results are compared to the nominal ISO-8529 AmBe spectrum (black triangle).}
\end{figure}

Since a mono-energetic neutron source does not exist, the detector response function was obtained from the simulation with GEANT4 through the comparison of the unfolded AmBe neutron spectrum with the measured one. The back-scattering from the room was taken into account.

The unfolding program based on \cref{eq:SAND-II} was developed in C language. The measured nuclear recoil spectrum in \cref{fig:AmBeRecoilSpectrum} was used as the input to unfold the neutron spectrum of the AmBe source. The standard neutron spectrum of ISO-8529 AmBe and a flat spectrum were both adopted as the initial value $\Phi^{0}$ for the first iteration. The unfolded results are shown in \cref{fig:AmBeUnfoldingSpectrum}, indicating that most of the structures of the neutron spectrum were successfully reconstructed. The residual discrepancies can be attributed to the differences in source constructions between the ISO-8529 reference and the one used in the experiment. Furthermore, \cref{fig:AmBeUnfoldingSpectrum} shows that the final unfolded spectra are insensitive to the initial values. This character is of great importance for the neutron spectrum unfolding at CJPL, because a prior information on the spectrum is rarely known.

\section{Data analysis}
\label{sec:data analysis}

\subsection{Neutron flux measurements at CJPL}
\subsubsection{Hall A}
\label{sec:Hall A flux}

The detector was taking data in the Hall~A of CJPL (as shown in \cref{fig:CJPL_hall}) from October 2013 to January 2015 with live time of 356.28 days.

\begin{figure*}[t]
\centering
\begin{subfigure}{\columnwidth}
\includegraphics[width=\linewidth]{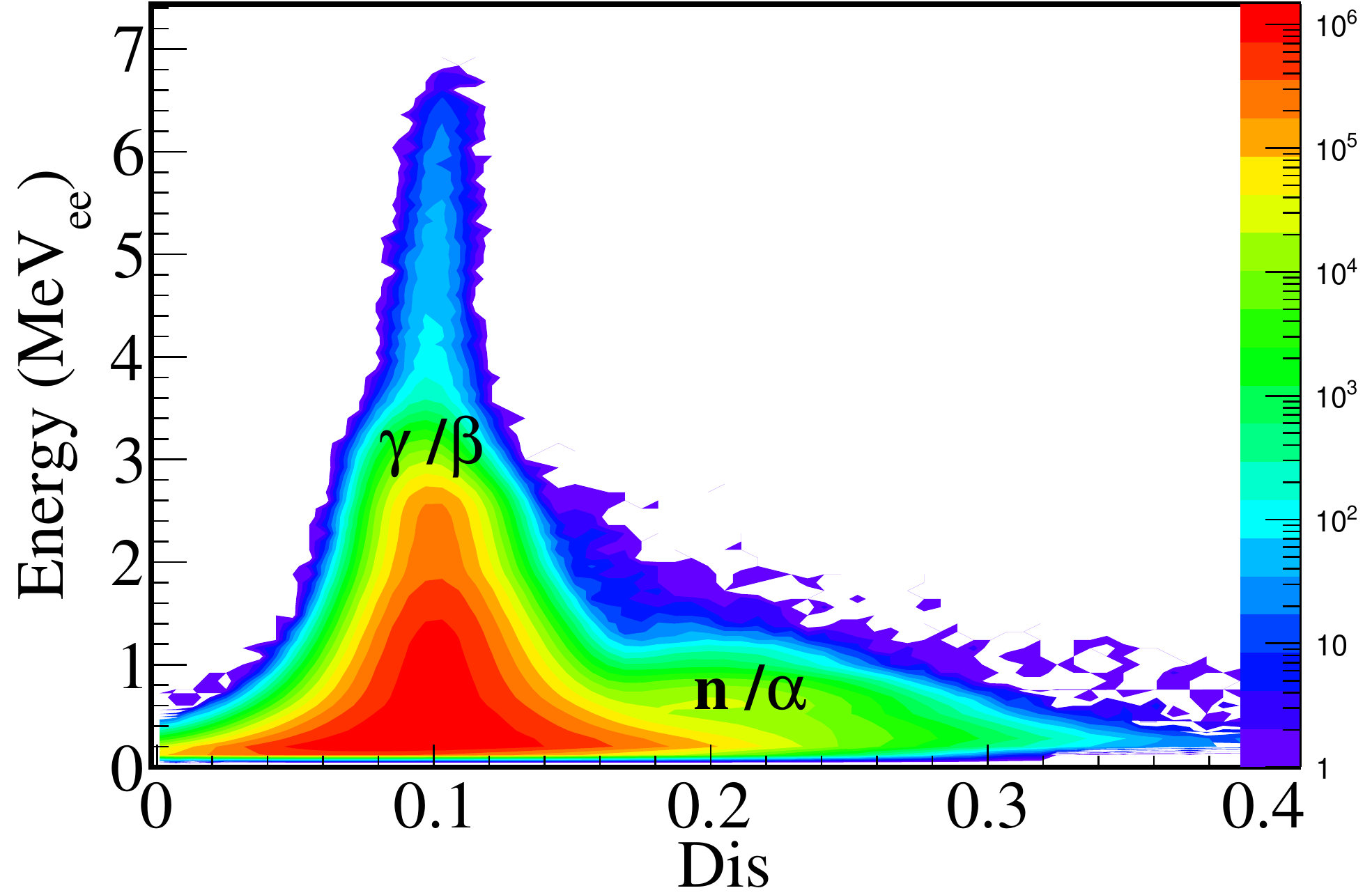}
\caption{\label{fig:CJPL_EnergyVsDisTotal}}
\end{subfigure}
\hfill
\begin{subfigure}{\columnwidth}
\includegraphics[width=\linewidth]{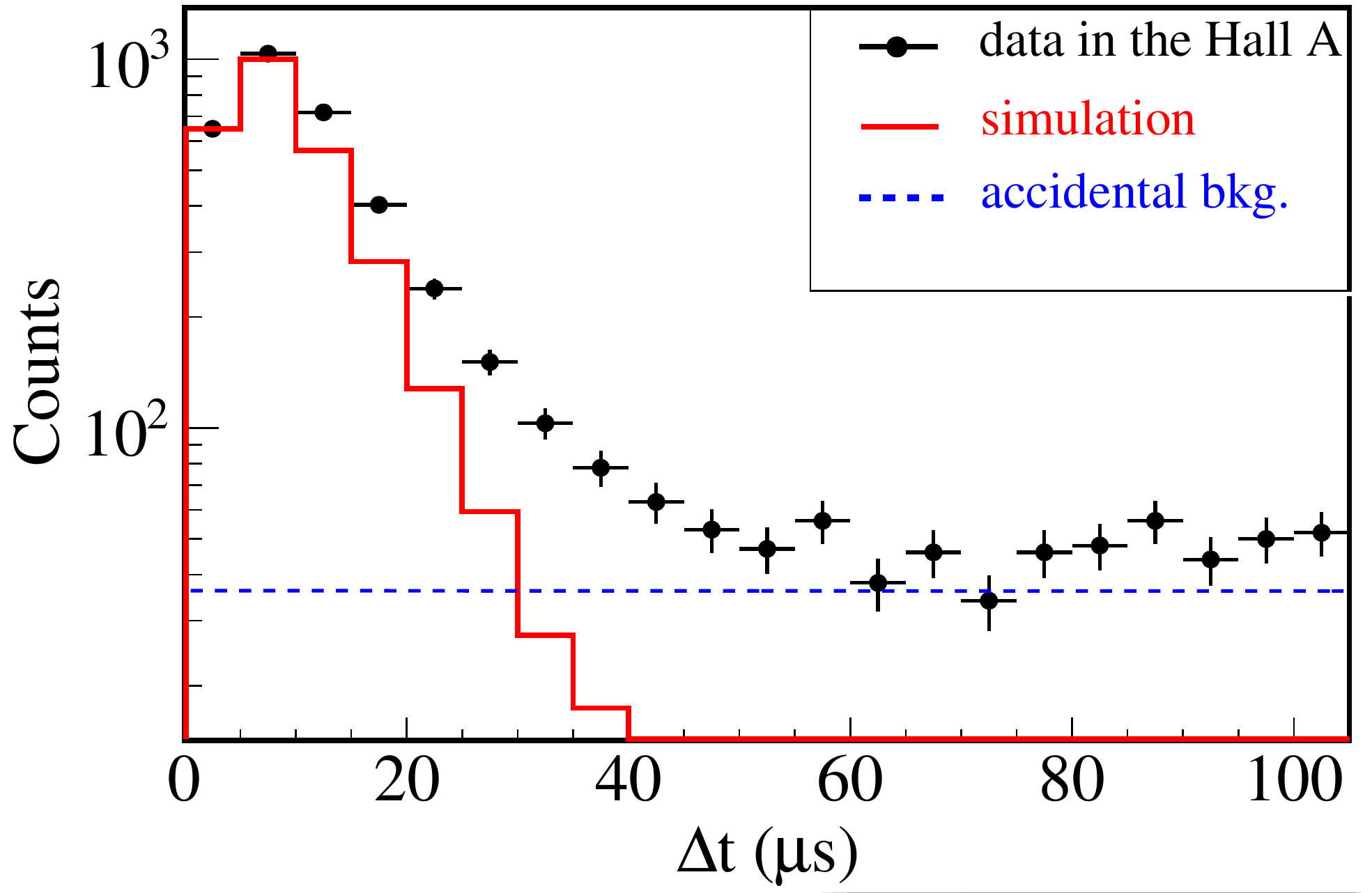}
\caption{\label{fig:CJPL_TimeInterval}}
\end{subfigure}
\begin{subfigure}{\columnwidth}
\includegraphics[width=\linewidth]{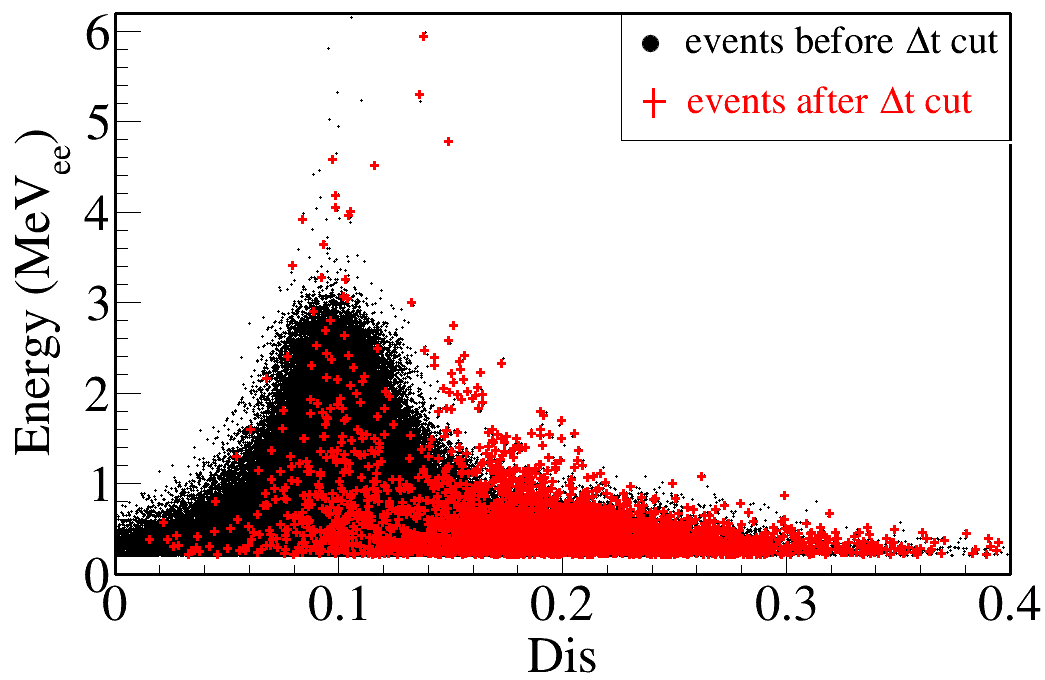}
\caption{\label{fig:CJPL_EnergyVsDisNeutron}}
\end{subfigure}
\hfill
\begin{subfigure}{\columnwidth}
\includegraphics[width=\linewidth]{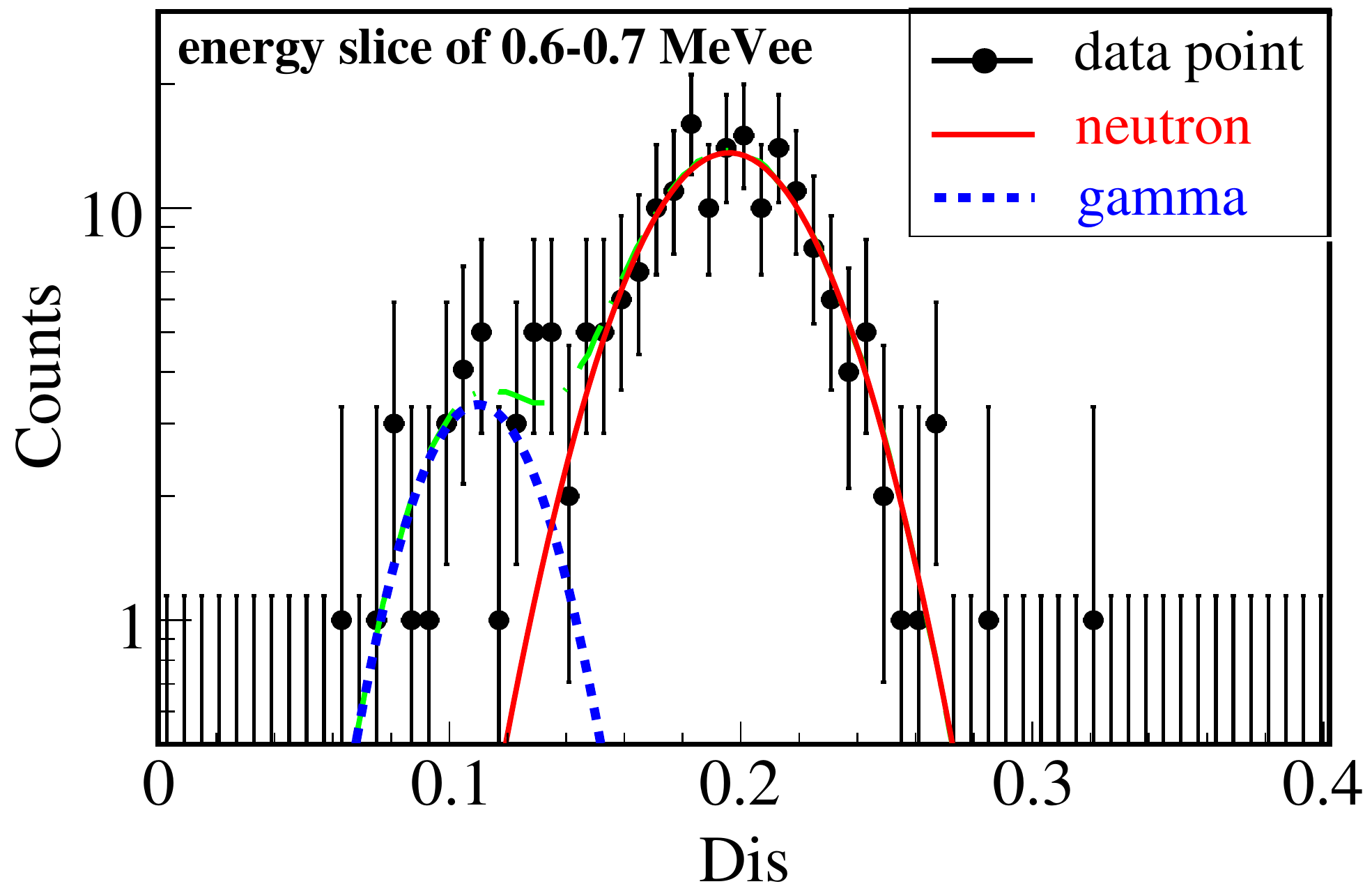}
\caption{\label{fig:energy_slice}}
\end{subfigure}
\caption{\label{fig:CJPL_selection}(color online) The selection procedure of nuclear recoils from the background data in CJPL Hall~A: (a) The discrimination factor ($Dis$) distribution at different energies for neutrons (n), $\upalpha$-particles, $\upgamma$-rays, and electrons ($\upbeta$); (b) The time interval distribution of the prompt-delayed signals; (c) Scatter plot of energy vs $Dis$ where the black points show the selected prompt events after applying the energy condition for delayed $\upgamma$-rays, while the red crosses show the selected prompt events after applying the coincidence time window; (d) Result of an unbinned maximum-likelihood fit of double-Gaussian to the events after selections in the 0.6\,--\,0.7\,MeV$_{ee}$ bin as illustration.}
\end{figure*}

\begin{table*}
\centering
\caption{\label{tab:selection} The selection procedure of prompt nuclear recoils. Listed are the individual ($\lambda\%$) and cumulative ($\Pi\lambda\%$) data survival fraction.}
\begin{tabular*}{\textwidth}{@{\extracolsep{\fill}}lccc}
\toprule
		    selection procedures    &  AmBe                & CJPL Hall A         & CJPL PE room\\
			\hline
			raw counts	
		                 &  $1.19\times 10^{6}$ & $8.59\times 10^{8}$ & $1.33\times 10^{8}$          \\
			\multicolumn{4}{l}{delayed $\upgamma$-rays$>$3 MeV}	
							 \\
			\multicolumn{1}{c}{$\lambda\%\ [\Pi\lambda\%]$ }
						 &  11.10 [11.10]        & 0.0446 [0.0446]      & 0.075 [0.075]                 \\
			\multicolumn{4}{l}{\tabincell{l}{prompt nuclear recoils$>$0.23 MeV$_{ee}^{\dagger}$}}	
							 \\
			\multicolumn{1}{c}{$\lambda\%\ [\Pi\lambda\%]$ }
						 &  80.25 [8.90]         & 85.42 [0.038]        & 73.14 [0.055]                  \\
			\multicolumn{4}{l}{prompt-delayed time window (2-30 $\upmu$s)}
							 \\
			\multicolumn{1}{c}{$\lambda\%\ [\Pi\lambda\%]$ }
						 &  22.81 [2.03]        & 0.938 [0.00036]      & 0.088 [0.000048]               \\				 
			\hline
			coincident $\upgamma$-background (bkg.)        & $1462\pm106$   &  $345\pm33$    & $19.4\pm5.7$   \\
			$\upgamma$-bkg. subtracted nuclear recoils (N) & $23159\pm322$  &  $2682\pm61$   & $44.1\pm7.4$   \\
			live time (t)                                  & 2697 s         & 356.28 d       & 173.55 d       \\
			detection efficiency ($\epsilon$\%)            & $11.13\pm0.42$ & $11.13\pm0.42$ & $11.13\pm0.42$ \\	
			total surface of the detector (cm$^{2}$)       & 5184 cm$^{2}$  & 5184 cm$^{2}$  & 5184 cm$^{2}$                           \\
			raw neutron flux at detector (cm$^{-2}$s$^{-1}$) &  ------     & $(1.51\pm0.03)\times10^{-7}$ 
			                                                                & $(5.1\pm0.9)\times10^{-9}$      \\
\bottomrule
\end{tabular*}
    \flushleft{\footnotesize $^\dagger$ Equivalent to 1 MeV neutron deposit energy}
\end{table*}

\cref{fig:CJPL_selection} illustrates the selection of the prompt nuclear recoils from the measured background data in CJPL Hall~A. 
The same procedure as described in \Cref{sec:efficiency} was applied. The black points in \cref{fig:CJPL_EnergyVsDisNeutron} show the selected prompt events after applying the energy condition for the delayed $\upgamma$-rays, while the red crosses show the selected prompt events after applying the coincidence time window. The raw spectrum of selected events (red crosses in \cref{fig:CJPL_EnergyVsDisNeutron}) is shown in \cref{fig:CJPL Hall recoil spectrum} (labeled as `signal+bkg.').

As in the neutron source measurement, there are residual $\upgamma$-ray background events. These events were rejected by using an unbinned maximum-likelihood fit to the $Dis$ with two Gaussian functions for each energy bin. \cref{fig:energy_slice} illustrates the fitting result for the energy bin of 0.6\,--\,0.7\,MeV$_{ee}$. 
The fit results provide the number of nuclear recoils and coincident $\upgamma$-ray background with their statistical uncertainties at each energy bin. The spectrum of the coincident $\upgamma$-ray background from the fit is also displayed in \cref{fig:CJPL Hall recoil spectrum}.

The neutron flux is calculated using the following formula:

\begin{equation}
\label{eq:flux}
F = \frac{N}{\epsilon\cdot t\cdot A}
\end{equation}
where $F$ stands for the neutron flux, $N = 2682\pm 61$ is the number of selected nuclear recoils (the number and its statistical uncertainty are given by the double-Gaussian fit), $\epsilon = (11.13\pm0.42)\%$ is the neutron detection efficiency under the selection conditions from the calibration data, $t = 356.28$ days is the live time of the data taking, $A = 5184 $\,cm$^{2}$ is the total surface area of the neutron detector (30\,cm in diameter and 40\,cm in length).
Accordingly, the measured neutron flux in CJPL Hall~A is $(1.51\pm0.03)\times10^{-7}$\,cm$^{-2}$s$^{-1}$ in the energy region from 1 to 10\,MeV. The upper limit comes from the dynamical range of the FADC.

\subsubsection{PE room}
After January 2015, the neutron detector was moved to the PE room where the high-purity germanium (HPGe) detectors of CDEX-1~\cite{CDEX-1} and CDEX-10~\cite{CDEX-10} are located, as illustrated in \cref{fig:CJPL_PE}. 
The neutron detector was on the top of the lead shield of CDEX-1B. 
The PE room is designed to shield against environmental neutrons, as well as $\upgamma$-ray background from the rock. 

\begin{figure*}[t]
\begin{subfigure}{\columnwidth}
\centering
\includegraphics[width=\linewidth]{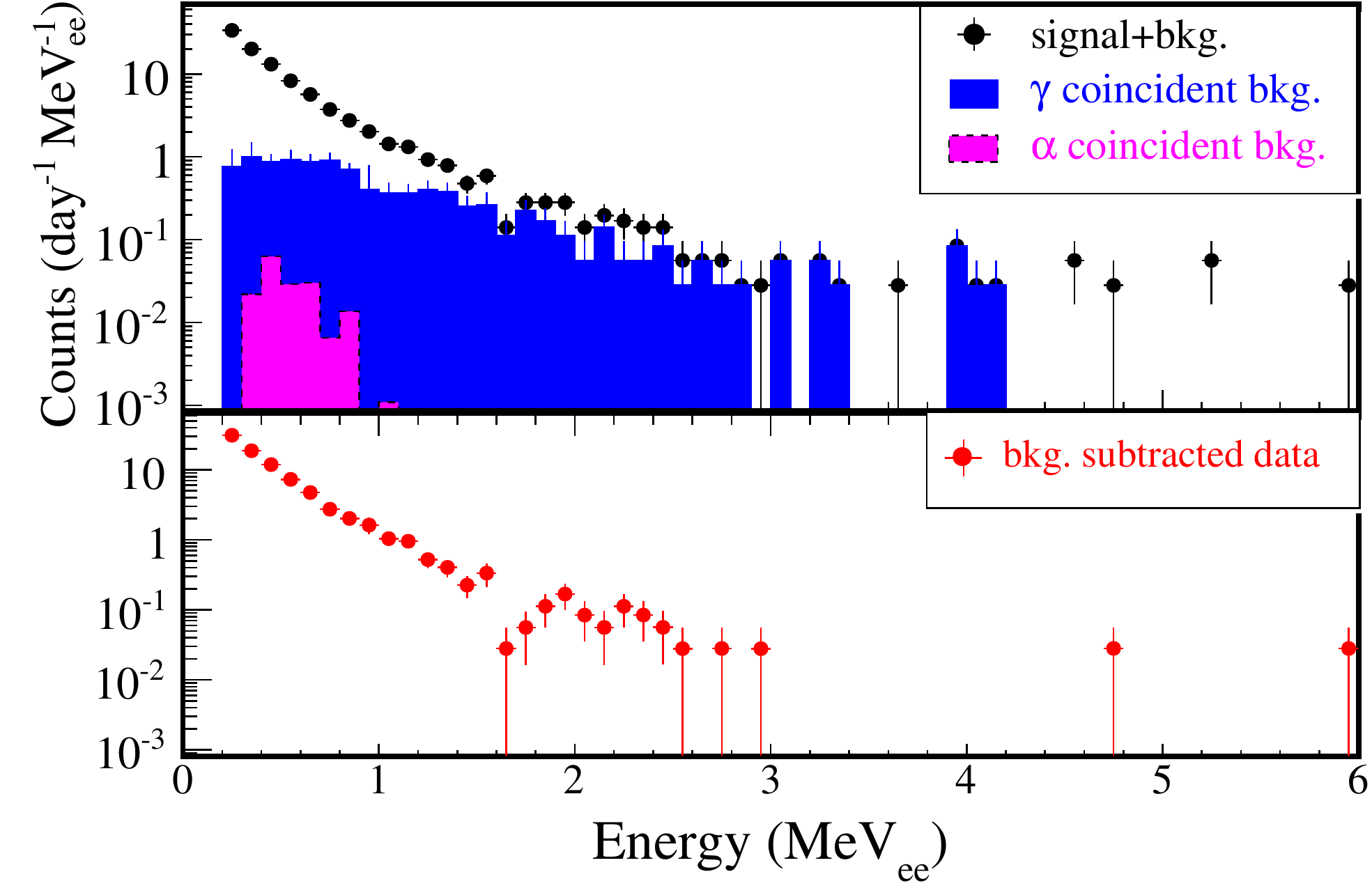}
\caption{\label{fig:CJPL Hall recoil spectrum}}
\end{subfigure}
\hfill
\begin{subfigure}{\columnwidth}
\centering
\includegraphics[width=\linewidth]{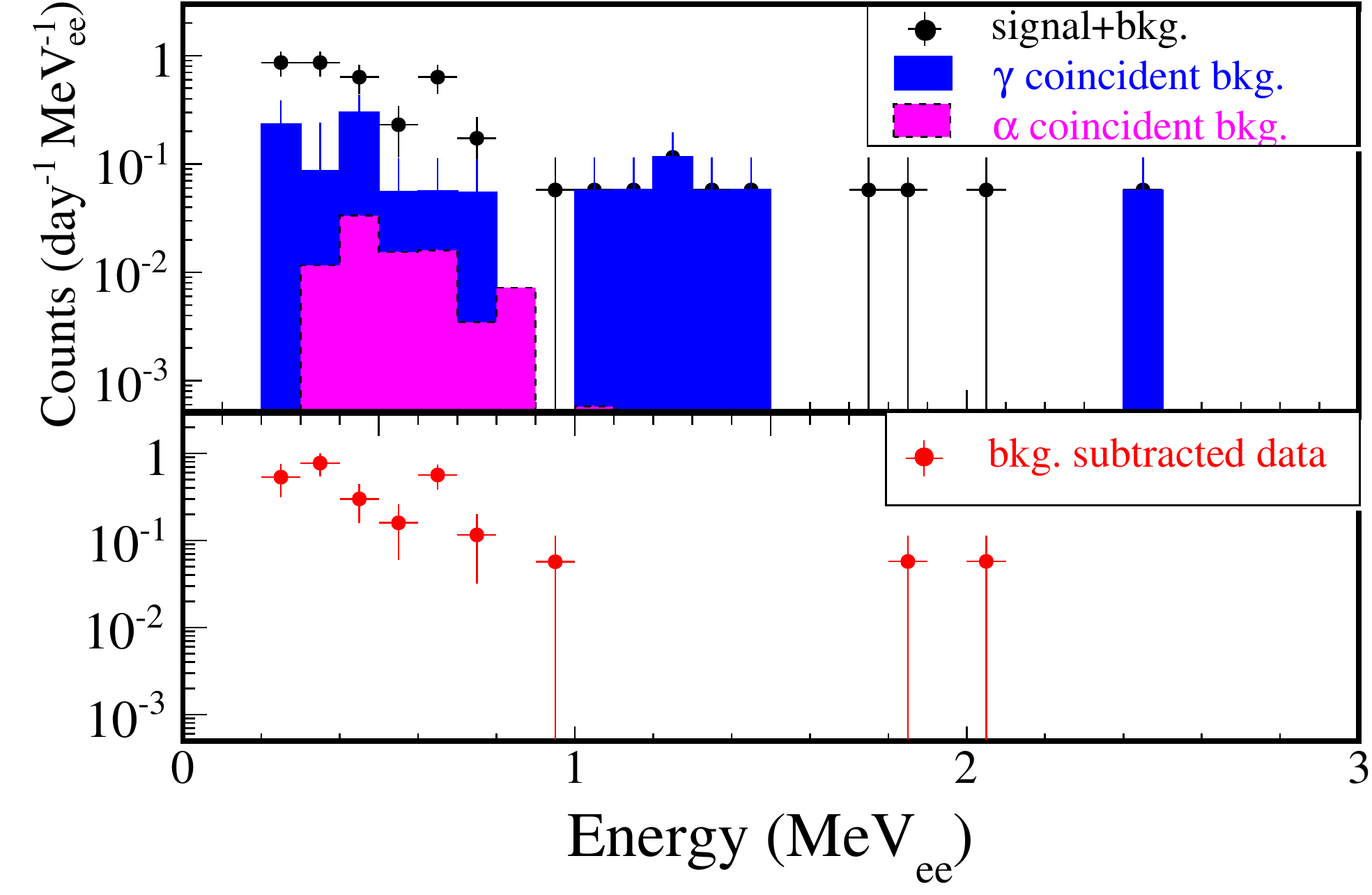}
\caption{\label{fig:CJPL PE room recoil spectrum}}
\end{subfigure}
\caption{\label{fig:CJPL_RecoilSpectrum}(color online) The raw spectra (signal+bkg.), $\upgamma$-ray coincident background ($\upgamma$~coincident~bkg.) and $\alpha$ random coincident background ($\upalpha$~coincident~bkg.) are displayed at the top panels, leading to the final background subtracted nuclear recoil spectra of CJPL (bkg.~subtracted~data) at the bottom panels (only statistical uncertainties are shown). The $\upgamma$-ray coincident background was subtracted by the double-Gaussian fit. (a) The measured nuclear recoil spectra in the Hall~A; (b) The measured nuclear recoil spectra in the PE room. (see text for details)}
\end{figure*}

After using the same procedure as for the Hall~A, the raw spectrum (labeled as `signal+bkg.') and the coincident $\upgamma$-ray background spectrum are shown in \cref{fig:CJPL PE room recoil spectrum}.
Based on the live time of 173.55 days, $N = 44.1\pm 7.4$ nuclear recoils are obtained. 
Therefore, the measured neutron flux in the PE room of CJPL is $(5.1\pm 0.9) \times 10^{-9}$\,cm$^{-2}$s$^{-1}$ in the energy range of 1\,--\,10\,MeV. 

The selection procedure of nuclear recoils and the measured neutron fluxes are summarized in \Cref{tab:selection} both for CJPL Hall~A and PE room, as well as the calibration data with the AmBe neutron source.

\begin{table*}[t]
\centering
\caption{\label{tab:background} The $\alpha$-particle background estimation for the data in the Hall~A and in the PE room.}
     \begin{tabular*}{\textwidth}{@{\extracolsep{\fill}}lcc}
     \toprule
		        background               & CJPL Hall A           & CJPL PE room \\
			\hline				                         			
			\tabincell{c}{$\alpha$ coincidence with $\upgamma$ ($>$3 MeV) (accidental)} 
			                             & $5.88\pm0.03$         & $1.52\pm0.01$ \\

			\tabincell{c}{$\alpha$ coincidence with [$(\alpha+\upgamma)>$3 MeV] (physical cascade decay)} 
			                             & $<4.3\times10^{-4}$   & $<3.5\times10^{-5}$ \\	

			\tabincell{c}{$\alpha$ coincidence with [$(\alpha+\upgamma)>$3 MeV] (accidental)} 
			                             & $<1.5\times10^{-4}$   & $<2.5\times10^{-5}$ \\
 			\hline
			\tabincell{c}{total $\alpha$-background (bkg.)} 
			                             & $5.88\pm0.03$         & $1.52\pm0.01$ \\
			\tabincell{l}{$\upgamma$-bkg. subtracted nuclear recoils}           
			                             & $2682\pm61$           & $44.1\pm7.4$ \\			                              

			\tabincell{c}{$\alpha$-bkg. subtracted nuclear recoils} 
			                             & $2676\pm61$           & $42.6\pm7.4$ \\		

			\tabincell{c}{live time (day)} 
			                             & 356.28                & 173.55  \\					                             	                             
			\tabincell{c}{final neutron flux at detector (cm$^{-2}$s$^{-1}$)} 
			                             & $(1.51\pm0.03)\times10^{-7}$ & $(4.9\pm0.9)\times10^{-9}$ \\	                             
	\bottomrule						
    \end{tabular*}
\end{table*}

\begin{table*}
\centering
\caption{\label{tab:neutron_flux} Fast neutron fluxes at different underground laboratories}
     \begin{tabular*}{\textwidth}{@{\extracolsep{\fill}}cccc}
     \toprule
		\tabincell{c}{Underground\\laboratory} &\tabincell{c}{Thickness of\\ rock overburden (m)} &\tabincell{c}{Fast neutron flux\\(cm$^{-2}$s$^{-1}$)} &\tabincell{c}{Energy range\\(MeV)} \\
			\hline
			CJPL PE room & 2400 &  $(4.9\pm0.9\,(stat.)\pm0.5\,(syst.))\times10^{-9} $ (this work)       &  1--10  \\
			CJPL Hall A  & 2400 &  $(1.51\pm0.03\,(stat.)\pm0.10\,(syst.))\times10^{-7} $ (this work)     &  1--10  \\
			LSM          & 1780 &  $(4.0\ \pm\ 1.0)\times10^{-6}$ \cite{Modane}  \qquad\qquad &  2-- 6 \\
			Gran Sasso   & 1400 &  $(4.20\pm0.08)\times10^{-7}$ \cite{GranSasso} \qquad\qquad &  1--10  \\
			Boulby       & 1070 & $(1.72\pm0.61\,(stat.)\pm0.38\,(syst.))\times10^{-6}$ \cite{Boulby}   &  $>$0.5  \\
			Canfranc     & 800  & $(4.10\pm0.08)\times10^{-7}$ \cite{Canfranc}  \qquad\qquad &  1--10    \\
			YangYang     & 700  & $(4.17\pm0.90)\times10^{-6}$ \cite{YangYang}  \qquad\qquad &  1--10    \\
			CPL          & 350  & $(3.00\pm0.02\,(stat.)\pm0.05\,(syst.)\times10^{-5}$ \cite{CPL}      &  1.5--6    \\			
    \bottomrule						
    \end{tabular*}
\end{table*}

\subsection{Discussion of $\alpha$ background}
Despite the unique experimental signatures and the strict selection criteria, there may be residual $\alpha$ background present. We discuss their origin and estimate their contribution in this section.

\subsubsection{$\alpha$ random coincidence with $\gamma$-rays}
\label{sec:alpha-bg}
When the \textit{in-situ} $\alpha$-particles from U/Th series inside the Gd-LS are in coincidence with $\upgamma$-ray events (E$_{\upgamma}>$3\,MeV) during the time window of (2, 30)\,$\upmu$s , these $\alpha$-particles can pass the selection criteria because of the disability to distinguish the $\alpha$-particles from neutrons by the PSD method.
 
The total rate of the intrinsic $\alpha$-particles is estimated to be $(0.548\pm0.002)$\,s$^{-1}$ based on the contamination levels of U/Th in the Gd-LS, assuming secular equilibrium and considering all the isotopes in the decay chains. The intrinsic contaminations of U/Th of the Gd-LS are discussed in a forthcoming publication~\cite{intrinsic contamination}. The rate of $\upgamma$-rays with energy larger than 3\,MeV is $(1.245\pm0.002)\times10^{-2}$\,s$^{-1}$ and $(6.63\pm0.02)\times10^{-3}$\,s$^{-1}$ in the Hall A and the PE room, respectively. The random coincident background from $\alpha$-particles is therefore estimated to be $(5.88\pm0.03)$ events out of $(2682\pm61)$ nuclear recoil events in the Hall~A, and $(1.52\pm0.01)$ events out of $(44.1\pm7.4)$ nuclear recoil events in the PE room, respectively.

The spectra of $\alpha$ background are also estimated in \cref{fig:CJPL_RecoilSpectrum} (labeled as `$\alpha$~coincident~bkg.') based on the quenching of $\alpha$-particles in the Gd-LS~\cite{intrinsic contamination}.

\subsubsection{$\alpha$ coincidence with ($\alpha$ + $\gamma$-rays)}
An $\alpha$-particle in coincidence with a $\upgamma$-ray within 40\,ns can be misidentified as a high energy $\upgamma$-ray event. Having an additional $\alpha$-particle at 2\,--\,30\,$\upmu$s before this one would be selected as a neutron event. This background has two main sources: one is from the $\alpha$-$\alpha$ cascade decay of $^{219}$Rn$\to ^{215}$Po$\to ^{211}$Pb, and the other is from $\alpha$-$\alpha$ accidental coincidence.

The $\alpha$-$\alpha$ cascade and the accidental coincidence backgrounds are estimated to be $<4.3\times10^{-4}$ and $<1.5\times10^{-4}$ events in the Hall A, while $<3.5\times10^{-5}$ and $<2.5\times10^{-5}$ events in the PE room respectively. They are therefore negligible compared to the $\alpha$ accidental background discussed in \Cref{sec:alpha-bg}. 

The results of all $\alpha$-particle backgrounds discussed in this section are listed in \Cref{tab:background}.

\section{Results}
\subsection{Fast neutron fluxes at CJPL}
The systematic uncertainties are mainly due to the $\upgamma$-ray energy calibration caused by the light yield shift of the Gd-LS during the measurement (3.3\% for the Hall~A and 7.4\% for the PE room, respectively), 3.8\% from the efficiency calibration by AmBe source, and 4.3\% from the double-Gaussian fit (estimated by fitting the $^{60}$Co calibration data).

After taking into account the estimated $\alpha$-particle background in \Cref{tab:background} and the systematic uncertainties, the final neutron fluxes in the 1 to 10\,MeV region are $(1.51\pm0.03\,(stat.)\pm0.10\,(syst.))\times10^{-7}$\,cm$^{-2}$s$^{-1}$ and $(4.9\pm0.9\,(stat.)\pm0.5\,(syst.) )\times10^{-9}$\,cm$^{-2}$s$^{-1}$ for the Hall~A and the PE room, respectively. 
Thus the $\alpha$-particle backgrounds can be neglected for the measurement of fast neutron background in the Hall~A. \cref{fig:CJPL_RecoilSpectrum} displays the $\upgamma$ and $\alpha$ backgrounds subtracted nuclear recoil spectra both for the Hall~A and PE room.

\Cref{tab:neutron_flux} shows that, CJPL Hall~A has the same level of environmental fast neutron background as Gran Sasso and Canfranc, but much lower than the other underground laboratories due to the low levels of U/Th contamination in the rock. 

The measured neutron flux inside the PE room is reduced by more than one order of magnitude compared to the Hall~A. 
This attenuation can be attributed to the one meter thick PE. The residual neutron flux is expected to originate from the $(\alpha, n)$ processes from the natural radioactivity of experimental hardware. 

The sensitivities of the detector to fast neutron flux can be estimated by the uncertainties of the $\upgamma$-ray and $\alpha$-particle coincident backgrounds. After considering the systematic uncertainty of the double-Gaussian fit,
the detection limits of fast neutron background at 90\% confidence level are $2.6\times10^{-9}$ and $8.5\times10^{-10}$\,cm$^{-2}$s$^{-1}$ in the Hall~A and the PE room, respectively. The sensitivities are mainly limited by the uncertainties of the $\upgamma$-ray coincident background.

\subsection{Fast neutron spectra at CJPL}
The same method discussed in \Cref{sec:AmBe spectrum unfolding} was used to reconstruct the spectra of fast neutron background both in the Hall A and the PE room. The detector response function was derived from simulation with mono-energetic neutron sources from 1 to 12\,MeV with 0.1\,MeV step, placed at the surface of a cuboid with dimension of 2\,$\times$\,1\,$\times$\,1\,m$^3$. The cuboid is fully enclosing the lead castle of the neutron detector. The angular distribution was set to be isotropic. 

\begin{figure}
\includegraphics[width=\linewidth]{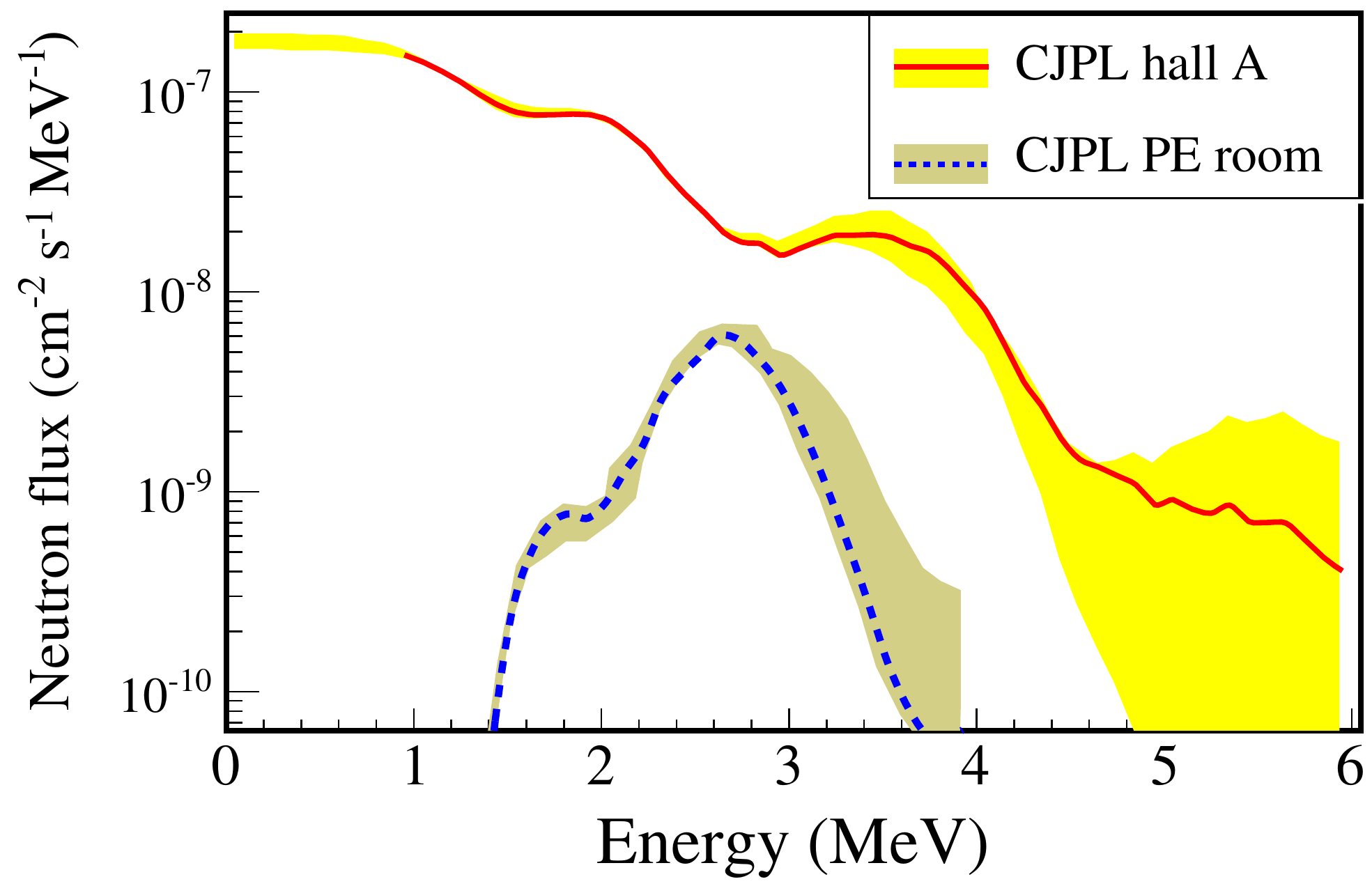}
\caption{\label{fig:CJPLNeutronSpectrum}(color online) The reconstructed fast neutron spectra both in the Hall~A and the PE room at CJPL. The areas are the $\pm\sigma$ uncertainty bands of the reconstructed spectra.}
\end{figure}

The background subtracted nuclear recoil spectra in \cref{fig:CJPL_RecoilSpectrum} were used as the input for the unfolding program, while a flat spectrum was fed as the initial value for the iterations. The resulting fast neutron spectra and their 1$\sigma$ uncertainty band are displayed in \cref{fig:CJPLNeutronSpectrum}. The shape of the reconstructed fast neutron spectrum in the Hall~A is similar as the spectrum estimated at LSM~\cite{Modane}.

\section{Conclusion}
\label{sec:conclusion}
We presented the results of the measurement of fast neutron background with a 0.5\% gadolinium doped liquid scintillator in the Hall A and the one meter thick PE room at CJPL.  
The measured neutron fluxes in the 1\,--\,10\,MeV energy range are $(1.51\pm0.03\,(stat.)\pm0.10\,(syst.))\times10^{-7}$\,cm$^{-2}$s$^{-1}$ and $(4.9\pm0.9\,(stat.)\pm0.5\,(syst.) )\times10^{-9}$\,cm$^{-2}$s$^{-1}$ for CJPL Hall A and the PE room, respectively.
The corresponding fast neutron spectra in the Hall A and the PE room were also reconstructed with the help of GEANT4 simulation.

A factor of thirty improvement of the fast neutron background in the PE room has been achieved. However, the shield of the one meter polyethylene is expected to bring six orders of magnitude reduction for 3\,MeV neutrons from simulation. This suggests that the neutron flux measured in the PE room may have contributions from the materials of the experimental setup and/or from the broken-chain of the radon in the room. 
The projected sensitivities of neutron flux for the prompt-delayed coincidence method are estimated to be $2.6\times10^{-9}$ and $8.5\times10^{-10}$\,cm$^{-2}$s$^{-1}$ at 90\% confidence level in the Hall~A and in the PE room, respectively. 

Improved measurements with the detector at a threshold of $O$(100\,eV) and covered with additional PE plates are being pursued. The quantitative understanding of the various neutron sources can be expected through the comparisons of flux and spectral shape under different configurations.


\section*{Acknowledgments}

This work was supported by the National Natural Science Foundation of China (Contracts No. 11275134, No. 11475117, No. 11475099), 
National Basic Research Program of China (973 Program) (Contract No. 2010CB833006) 
and Academia Sinica Principal Investigator 2011-2015 
and Contract 104-2112-M-001-038-MY3 from the Ministry of Science and Technology, Taiwan.
Qiang Du is grateful to Max-Planck-Institut f\"ur Physik for hospitality where part of this work was carried out.




\nocite{*}
\bibliographystyle{elsarticle-num}
\bibliography{jos}

\begin{thebibliography}{00}


 \bibitem{dark-matter} J. L. Liu et al., \href{https://www.nature.com/nphys/journal/v13/n3/abs/nphys4039.html}{Nat. Phys. 13 (2017) 212–216}; C. Patrignani et al. (Particle Data Group), \href{http://iopscience.iop.org/article/10.1088/1674-1137/40/10/100001/meta}{Chin. Phys. C 40 (2016) 100001}, and references therein.
 \bibitem{proton-decay}K. Abe et al. (Super-Kamiokande Collaboration),
\href{http://journals.aps.org/prd/abstract/10.1103/PhysRevD.95.012004}{Phys. Rev. D 95 (2017) 012004}; C. Patrignani et al. (Particle Data Group), \href{http://iopscience.iop.org/article/10.1088/1674-1137/40/10/100001/meta}{Chin. Phys. C 40 (2016) 100001}, and references therein.
 \bibitem{neutrinoless}M. Agostini et al. (GERDA Collaboration), \href{https://www.nature.com/nature/journal/v544/n7648/full/nature21717.html}{Nature 544 (2017) 47–52}; C. Patrignani et al. (Particle Data Group), \href{http://iopscience.iop.org/article/10.1088/1674-1137/40/10/100001/meta}{Chin. Phys. C 40 (2016) 100001}, and references therein.
 \bibitem{neutrino} Q. R. Ahmad et al. (SNO Collaboration), \href{https://journals.aps.org/prl/abstract/10.1103/PhysRevLett.87.071301}{Phys. Rev. Lett. 87 (2001) 071301}; S. Fukuda et al. (Super-Kamiokande Collaboration), \href{https://doi.org/10.1103/PhysRevLett.86.5656}{Phys. Rev. Lett. 86 (2001) 5656}; C. Patrignani et al. (Particle Data Group), \href{http://iopscience.iop.org/article/10.1088/1674-1137/40/10/100001/meta}{Chin. Phys. C 40 (2016) 100001}, and references therein.
 \bibitem{CJPL}J. P. Cheng et al., \href{http://www.annualreviews.org/doi/abs/10.1146/annurev-nucl-102115-044842}{Annu. Rev. Nucl. Part. Sci. 67 (2017) 1}.
 \bibitem{Muon}Y. C. Wu et al., \href{http://iopscience.iop.org/article/10.1088/1674-1137/37/8/086001}{Chin. Phys. C 37 (2013) 086001}.
 \bibitem{Gamma}Z. Zeng et al., \href{http://link.springer.com/article/10.1007/s10967-014-3114-1}{J Radioanal. Nucl. Chem. 301 (2014) 443}.
 \bibitem{CDEX}K. J. Kang et al., \href{http://link.springer.com/article/10.1007/s11467-013-0349-1}{Front. Phys. 8 (2013) 412}; W. Zhao et al. (CDEX Collaboration), \href{http://journals.aps.org/prd/abstract/10.1103/PhysRevD.93.092003}{Phys. Rev. D 93 (2016) 092003}; S. K. Liu et al. (CDEX Collaboration), \href{https://journals.aps.org/prd/abstract/10.1103/PhysRevD.95.052006} {Phys. Rev. D 95 (2017) 052006}.
 \bibitem{PandaX}M. J. Xiao et al., \href{http://link.springer.com/article/10.1007/s11433-014-5598-7}{Sci. China Phys. Mech. Astron. 57 (2014) 2024}; A. Tan et al. (PandaX-II Collaboration), \href{https://journals.aps.org/prl/abstract/10.1103/PhysRevLett.117.121303}{Phys. Rev. Lett. 117 (2016) 121303}; C. Fu et al. (PandaX-II Collaboration), \href{https://journals.aps.org/prl/abstract/10.1103/PhysRevLett.118.071301}{Phys. Rev. Lett. 118 (2017) 071301}.
 \bibitem{thermal neutron} Z. M. Zeng et al., \href{https://doi.org/10.1016/j.nima.2015.09.043}{Nucl. Instrum. Meth. A 804 (2015) 108-112}.
 \bibitem{thermal neutron PE} Z. M. Zeng et al., \href{https://doi.org/10.1016/j.nima.2017.04.009}{Nucl. Instrum. Meth. A 866 (2017) 242-247}.
 \bibitem{Hu Q} Q. D. Hu et al., \href{http://www.sciencedirect.com/science/article/pii/S0168900217304023}{Nucl. Instrum. Meth. A 859 (2017) 37-40}.
 \bibitem{EJ-335} \href{http://www.eljentechnology.com/products/liquid-scintillators/ej-331-ej-335}{www.eljentechnology.com/products/liquid-scintillators/ej-331-ej-335}.
 \bibitem{Gd-isotopes}S. F. Mughabghab, \href{https://www-nds.iaea.org/relnsd/NdsEnsdf/neutroncs.html}{Atlas of Neutron Resonances Elsevier Science, April 17 (2006)}.
 \bibitem{Gd-Q}\href{https://www-nds.iaea.org/exfor/servlet/E4sGetIntSection?SectID=6001683&req=1574&e4up=0}{ENDF/B-VIII.b4(Gd155)}, \href{https://www-nds.iaea.org/exfor/servlet/E4sGetIntSection?SectID=6002300&req=1574&e4up=0}{ENDF/B-VIII.b4(Gd157)}.
 \bibitem{Boulby}E. Tziaferi et al., \href{http://www.sciencedirect.com/science/article/pii/S0927650506001873}{Astropart. Phys. 27 (2007) 326}.
 \bibitem{Aberdeen Tunnel} S. C. Blyth et al. (Aberdeen Tunnel Experiment Collaboration), \href{https://journals.aps.org/prd/abstract/10.1103/PhysRevD.93.072005}{Phys. Rev. D 94 (2016) 099906}.
 \bibitem{Double Chooz}F. Ardellier et al. (Double Chooz Collaboration), \href{https://arxiv.org/abs/hep-ex/0606025}{ 	arXiv:hep-ex/0606025}.
 \bibitem{RENO} J. K. Ahn et al. (RENO Collaboration), \href{https://arxiv.org/abs/1003.1391}{arXiv:1003.1391 [hep-ex]}.
 \bibitem{Daya Bay} F. P. An et al., \href{http://journals.aps.org/prl/abstract/10.1103/PhysRevLett.108.171803}{Phys. Rev. Lett. 108 (2012) 171803}.
 \bibitem{Mei_detector} C. Zhang et al., \href{http://www.sciencedirect.com/science/article/pii/S0168900213009881}{Nucl. Instr. and Meth. A 729 (2013) 138} and references therein.
 \bibitem{SAND}S. L. Wang et al., \href{http://iopscience.iop.org/article/10.1088/1674-1137/33/5/012/meta}{Chin. Phys. C 33 (2009) 378}; W. N. McElroy et al., U.S. Air Force Weapons Laboratory Report AFWL-TR-67-41 (1967); H. Sekimoto and N. Yamamuro, Nucl. Sci. Eng. 80 (1982) 101; J. T. Routti and J. V. Sandberg, Rad. Prot. Dosim. 10 (1985) 103.
 \bibitem{Shukui} S. K. Liu et al., \href{http://iopscience.iop.org/article/10.1088/1674-1137/39/4/046002/meta}{Chin. Phys. C 39 (2015) 046002}.
 \bibitem{Birks} J. B. Birks, \href{http://www.sciencedirect.com/science/book/9780080104720}{The Theory and Practice of Scintillation  Counting (Pergamon, New York, 1964)}.
 \bibitem{TRIM}\href{http://www.srim.org/}{http://www.srim.org/}.
 \bibitem{TypicalQuenching}M. Anghinolfi et al., \href{http://www.sciencedirect.com/science/article/pii/0029554X79902738}{Nucl. Instr. and Meth. A 165 (1979) 217}.
 \bibitem{CDEX-1} W. Zhao, et al. (CDEX Collaboration), \href{http://journals.aps.org/prd/abstract/10.1103/PhysRevD.88.052004}{Phys. Rev. D 88 (2013) 0520045}.
 \bibitem{CDEX-10}S. K. Liu and Q. Yue, \href{http://www.worldscientific.com/doi/abs/10.1142/S0217751X15450074}{Int. J. Mod. Phys. A 30 (2015) 1545007}.
 \bibitem{intrinsic contamination} Q. Du et at., Quenching of gadolinium doped liquid scintillator measurement based on intrinsic contamination of Uranium and Thorium, in preparation.
 \bibitem{Modane}V. Chazal et al., \href{http://www.sciencedirect.com/science/article/pii/S0927650598000127}{Astropart. Phys. 9 (1998) 163}.
 \bibitem{GranSasso}F. Arneodo et al., \href{https://inis.iaea.org/search/search.aspx?orig_q=RN:31027379}{Il Nuovo Cim. 112A (1999) 819}.
 \bibitem{Canfranc}D. Jordan et al., \href{http://www.sciencedirect.com/science/article/pii/S0927650512002046}{Astropart. Phys. 42 (2013) 1}.
 \bibitem{YangYang}H. Park et al., \href{http://www.sciencedirect.com/science/article/pii/S0969804313001589?np=y&npKey=b81d43a6167c373507a575a6c53e703fc7d08172ebd0b791709f299867591c31}{Applied Radiation and Isotopes 81 (2013) 302}.
 \bibitem{CPL}H.J. Kim et al., \href{http://www.sciencedirect.com/science/article/pii/S0927650503002433}{Astropart. Phys. 20 (2004) 549}.

 \end{thebibliography}



\end{document}